\documentclass[aps,pra,superscriptaddress,twocolumn,citeautoscript,10pt, rsi,reprint]{revtex4-1}
\usepackage{amsmath}
\usepackage{graphicx}
\usepackage{bm}
\usepackage{soul}
\usepackage{xcolor}
\usepackage{amsmath}
\usepackage{float}
\usepackage{comment}
\usepackage{graphicx}
\usepackage{dcolumn}
\usepackage[colorlinks=true,bookmarks=false,citecolor=blue,linkcolor=blue,hyperfootnotes=true,urlcolor=blue]{hyperref}

\usepackage{amsmath}
\usepackage{graphicx}
\usepackage{bm}
\usepackage{soul}
\usepackage{xcolor}
\usepackage{amsmath}
\usepackage{float}
\usepackage{comment}
\usepackage{graphicx}
\usepackage{dcolumn}

\def\GO{Ga$_2$O$_3$}
\def\BGO{$\beta$-Ga$_2$O$_3$}
\def\ALGO{(Al$_x$Ga$_{1-x}$)$_2$O$_3$}
\def\AO{Al$_2$O$_3$}
\def\TAO{$\theta$-Al$_2$O$_3$}

\begin{document}
\title{Role of carbon and hydrogen in limiting $n$-type doping of monoclinic \ALGO}
\author{Sai Mu}
\author{Mengen Wang}
\affiliation{Materials Department, University of California, Santa Barbara, California, 93106-5050, USA}
\author{Joel B. Varley}
\affiliation{Lawrence Livermore National Laboratory, Livermore, California 94450, USA}
\author{John L. Lyons}
\author{Darshana Wickramaratne}
\affiliation{Center for Computational Materials Science, United States Naval Research Laboratory, Washington, DC 20375, USA}
\author{Chris G. Van de Walle}\email{vandewalle@mrl.ucsb.edu}
\affiliation{Materials Department, University of California, Santa Barbara, California, 93106-5050, USA}

\begin{abstract}
We use hybrid density functional calculations to assess $n$-type doping in monoclinic {\ALGO} alloys.
We focus on silicon,
the most promising donor dopant, and study the structural properties, formation energies and charge-state transition levels of its various configurations.
We also explore the impact of
carbon and hydrogen, which are common impurities in metal-organic chemical vapor deposition (MOCVD).
In Ga$_2$O$_3$, Si$_\mathrm{Ga}$ is an effective shallow donor, but in Al$_2$O$_3$ Si$_\mathrm{Al}$ acts as a \textit{DX} center
with a ($+/-$) transition level in the band gap.
Interstitial hydrogen acts as a shallow donor in Ga$_2$O$_3$, but behaves as a compensating acceptor in $n$-type Al$_2$O$_3$.
Interpolation
indicates that Si is an effective donor in (Al$_x$Ga$_{1-x}$)$_2$O$_3$ up to 70\% Al, but it can be compensated by hydrogen already at 1\% Al.
We also assess the diffusivity of hydrogen and study complex formation.
Si$_\mathrm{cation}$--H complexes
have relatively low binding energies.
Substitutional carbon on a cation site acts as a shallow donor in Ga$_2$O$_3$, but can be stable in a negative charge state in (Al$_x$Ga$_{1-x}$)$_2$O$_3$ when
$x$$>$5\%.
Substitutional carbon on an oxygen site (C$_\mathrm{O}$) always acts as an acceptor in $n$-type (Al$_x$Ga$_{1-x}$)$_2$O$_3$,
but will incorporate only under relatively
oxygen-poor conditions.
C$_\mathrm{O}$--H complexes can actually incorporate more easily, explaining  observations of carbon-related compensation in Ga$_2$O$_3$ grown by MOCVD.
We also investigate C$_\mathrm{cation}$--H complexes, finding they have high binding energies and
act as compensating acceptors
when $x$$>$56\%; otherwise
the hydrogen just passivates the unintentional carbon donors.
C--H complex formation explains why MOCVD-grown Ga$_2$O$_3$ can exhibit record-low free-carrier concentrations, in spite of the  unavoidable incorporation of carbon.
Our study highlights that, while Si is in principle a suitable shallow donor in (Al$_x$Ga$_{1-x}$)$_2$O$_3$ alloys up to high Al compositions, control of unintentional impurities is essential to avoid compensation.

\end{abstract}
\maketitle

\section{\label{sec:level1}Introduction}

Monoclinic {\GO} ({\BGO}) is a wide-band-gap material (4.76---5.1 eV \cite{tippins1965optical,matsumoto1974absorption,sturm2016dipole,mock2017band}) with a high breakdown field (6-8 MV/cm) \cite{higashiwaki2012gallium}.
Despite its wide band gap, {\GO} can be controllably $n$-type doped;
together with the availability of high-quality yet low-cost substrates
these properties render {\BGO} highly promising for applications in high-power electronics and UV optoelectronics~\cite{suzuki2009enhancement,oshima2008vertical,alema2019solar}.

Effective control of the carrier concentration by doping with shallow donors is essential for device applications.
This typically requires that the donor impurity has a low ionization energy and that compensation can be avoided.
Si, Ge and Sn have been demonstrated to be effective shallow donors in {\GO} with modest ionization energies ($\le$ 80 meV) \cite{feng2019mocvd, zhang2019mocvd,son2016electronic,parisini2016analysis,ma2016intrinsic,higashiwaki2017state,moser2017ge,neal2018donors,orita2000deep,oishi2016conduction,higashiwaki2013depletion,varley2010oxygen,varley2020prospects}.  In addition to these dopants, first-principles calculations have identified interstitial hydrogen (H$_i$) and carbon on a Ga site (C$_\text{Ga}$) as shallow donors~\cite{varley2010oxygen,lyons2014carbon}.
Compensation of the shallow donors can occur due to native defects \cite{varley2011hydrogenated,ingebrigtsen2019impact} or impurities that act as acceptors, or due to self-compensation if the dopant can occur in different configurations.
One form of self-compensation involves the formation of substitutional species on either cation or anion sites.
For example, carbon in {\GO} can occur either on the Ga site (C$_\text{Ga}$), acting as a donor, or on the O cite (C$_\text{O}$), acting as an acceptor~\cite{lyons2014carbon}.
Self-compensation can also occur by formation of so-called \textit{DX} centers, in which an impurity expected to act as a shallow donor exhibits a large lattice relaxation and
becomes negatively charged by trapping electrons, thus effectively acting as a deep acceptor.
In {\GO}, Si, Ge, and Sn do not form \textit{DX} centers~\cite{varley2010oxygen,neal2018donors,von2020unusual}.
However, the likelihood of formation of a \textit{DX} center increases as the band gap increases, as is well known for AlGaAs~\cite{chadi1988theory} and AlGaN alloys~\cite{gordon2014hybrid}.

Alloying with Al raises the band gap of {\GO}~\cite{peelaers2018structural,*peelaers2019erratum,Varley2021_review}.
Heterojunctions of {\ALGO} and {\GO} exhibit a high-density two-dimensional electron gas, which is at the core of field-effect transistors \cite{zhang2018demonstration}.
Modulation doping is required for high mobility, raising the issue of whether $n$-type doping of {\ALGO} can be achieved.
Addressing this issue requires investigating whether dopants that are effective for {\GO} remain shallow donors in {\ALGO}, and whether compensation will occur.
In addition, recent experiments on {\ALGO} films grown by metal-organic chemical vapor deposition (MOCVD) indicated that control of doping might be challenging; it was found~\cite{zhao2020} that doping with Si failed to result in $n$-type doping below a threshold Si concentration.
Above the threshold, an abrupt enhancement of the carrier concentration
was observed.
Identifying the origin of the compensation in {\ALGO} alloys is therefore important for further improvements in electronic devices.

Varley \textit{et al.}~\cite{varley2020prospects} recently reported a first-principles study of \textit{DX}-center formation for a large number of candidate donor impurities in {\ALGO}.  They found that Si emerged as the best candidate, since it continues to act as a shallow donor up to high Al concentrations in the alloy.
In the present work we therefore focus on Si as the donor; we perform a more detailed study of its behavior as a \textit{DX} center, and investigate other sources of compensation and potential complex formation.

Our study, based on hybrid density functional theory (DFT), addresses Si impurities in monoclinic {\AO} (hereafter denoted as {\TAO}), and reveals a new configuration of the \textit{DX} state.
Compensation by native point defects has been previously addressed \cite{varley2011hydrogenated,ingebrigtsen2019impact}.
Here, we perform an in-depth study of compensation due to other impurities, particularly carbon and hydrogen.
Both of these impurities are readily incorporated during MOCVD growth.
Carbon on a cation site \cite{lyons2014carbon} and hydrogen \cite{varley2010oxygen} are both shallow donors in {\GO}, but possess ($0/-$) or ($+/-$) charge-state transition levels in the band gap in {\AO}.
By interpolating between the end compounds, {\GO} and {\AO}, we can estimate the position of these defect levels in {\ALGO} alloys, and determine the Al composition at which the onset of compensation occurs. [Note that we use the term ``defect'' to denote both native point defects and impurities.]
Our results show that Si is an effective donor in (Al$_x$Ga$_{1-x}$)$_2$O$_3$ up to 70\% Al, but it can be compensated by hydrogen already at 1\% Al, or by carbon on cation sites at 5\% Al.

Given that hydrogen is expected to be quite mobile, we also perform a study of its migration properties.
We further assess the possibility of forming hydrogen-related defect complex with Si and C.
Si$_\mathrm{cation}$--H complex formation is not a major concern, since the complexes have low binding energies.
Binding energies are much higher for C--H complexes; in fact, C--H behaves almost as a fixed entity, with properties very similar to a nitrogen impurity.
C$_\mathrm{cation}$--H complexes can act as compensating acceptors in (Al$_x$Ga$_{1-x}$)$_2$O$_3$ alloys, but only when the Al content exceeds 56\%; otherwise the complexes are mainly neutral and hydrogen passivates the unintentional carbon donors.
This  may explain why MOCVD-grown Ga$_2$O$_3$ can exhibit record-low free-carrier concentrations~\cite{alema2020}, in spite of the probably unavoidable incorporation of carbon.

Substitutional carbon on an oxygen site (C$_\mathrm{O}$), finally, always acts as an acceptor in $n$-type (Al$_x$Ga$_{1-x}$)$_2$O$_3$, irrespective of  Al concentration.  C$_\mathrm{O}$ will incorporate only under relatively cation-rich (oxygen-poor) conditions; we find that C$_\mathrm{O}$--H complexes can actually incorporate more easily, explaining experimental observations of carbon-related compensation in Ga$_2$O$_3$ grown by MOCVD~\cite{seryogin2020mocvd,alema2020h2o}.

The paper is organized as follows. In Sec.~\ref{method}, the calculational details, the definition of defect formation energies, and the physics of \textit{DX} centers are introduced. The main results for Si, C, and H and consequences for carrier compensation are presented in Secs.~\ref{siresult}, \ref{cresult}, and \ref{hresult}, respectively; complexes are discussed in Sec.~\ref{comp}.
Section~\ref{conc} concludes the paper.

\section{Methodology} \label{method}

\subsection{Computational details}\label{calc_detail}
\begin{figure}
\includegraphics[width=0.45\textwidth]{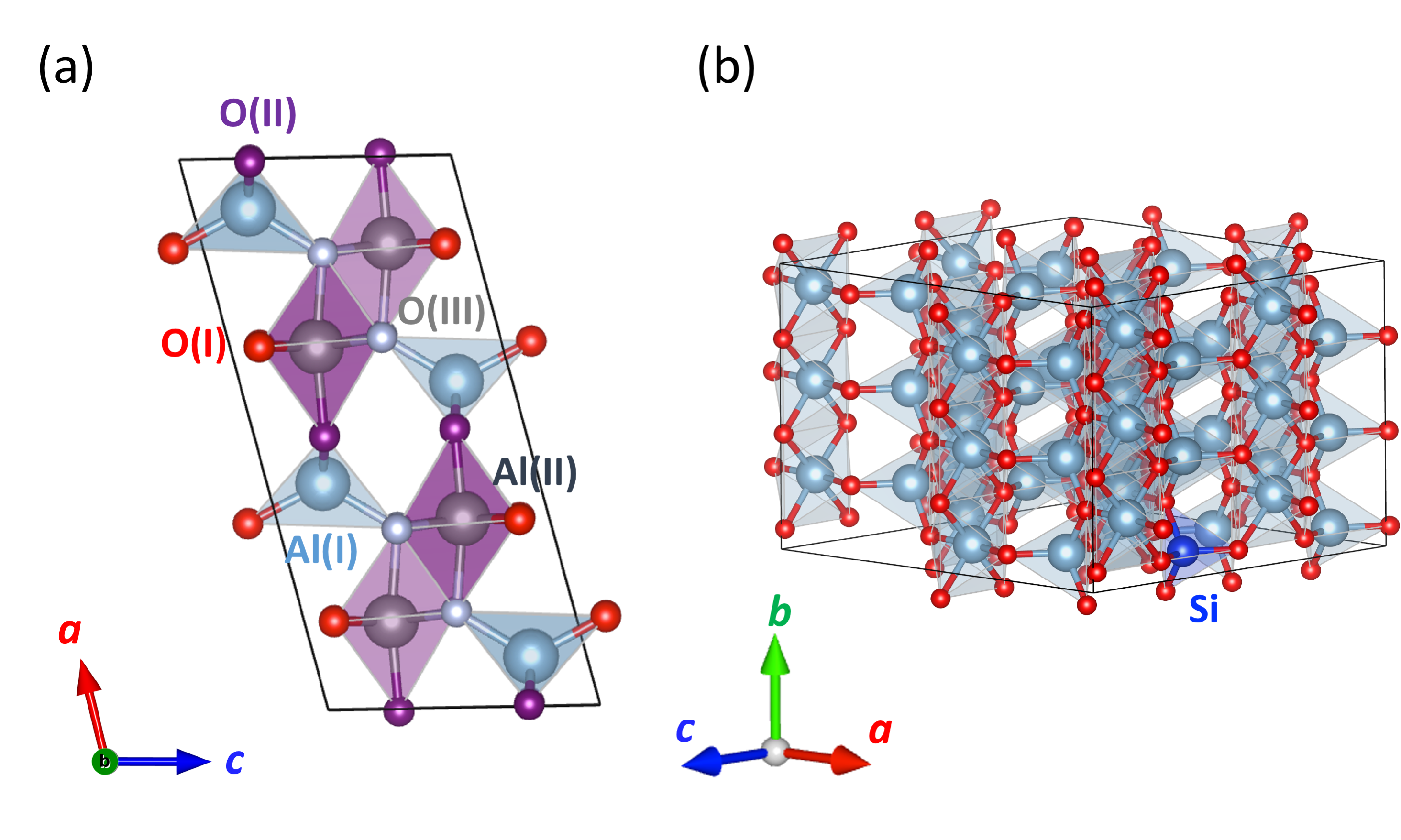}
\caption{\label{fig1:struct}
(a) Conventional cell of monoclinic {\AO}. Two inequivalent Al sites are labeled: tetrahedrally coordinated Al(I) in aqua and octahedrally coordinated Al(II) in purple.
For the O sites, threefold-coordinated O(I) are in red, threefold coordinated O(II) in purple, and fourfold- coordinated O(III) in gray.
(b) Supercell constructed by creating a $1\times3\times2$ multiple of the conventional monoclinic cell of {\TAO}, with one tetrahedrally-coordinated Al replaced by Si.  Light blue, red and dark blue spheres denote Al, O and Si atoms, respectively. Structural visualization was performed using VESTA \cite{momma2011vesta}.}
\end{figure}

We perform DFT calculations using the projector augmented wave method (PAW) \cite{Blochl1994} implemented in the Vienna \textit{Ab-initio} Simulation Package (VASP)  \cite{Kresse1993,Kresse1996}.
We focus on the monoclinic phase of {\GO} (denoted $\beta$-{\GO}) and {\AO} ($\theta$-{\AO}).  The structure, illustrated in Fig.~\ref{fig1:struct}(a), contains two types of cation sites: the tetrahedral site (denoted as I) and the octahedral site (denoted as II).
In addition, there are three types of O atoms: three-fold coordinated O(I) (on a shared corner of two edge-sharing AlO$_6$ octahedra and one AlO$_4$ tetrahedron), three-fold coordinated O(II) (on the shared corner of one AlO$_6$ octahedron and two AlO$_4$ tetrahedra), and four-fold coordinated O(III).

A plane-wave energy cutoff of 400 eV was employed and Brillouin-zone integration was carried out using the $\Gamma$-centered 4$\times$4$\times$4 k-point mesh for the primitive cell and 2$\times$2$\times$2 for the supercell.
The PAW potentials correspond to the valence-electron configurations 4$s^2$4$p^1$ for Ga, 3$s^2$3$p^1$ for Al, and 2$s^2$2$p^4$ for O.
To correctly describe the electronic structure as well as charge localization we use a hybrid functional, specifically the functional of Heyd, Scuseria, and Ernzerhof (HSE) \cite{heyd2003hybrid,*heyd2006erratum}, with a mixing parameter of $\alpha$=0.32.
This produces a band gap of 4.83 eV for {\BGO} and 7.41 eV for {\TAO}, in good agreement with the experimental band gaps of 4.80 for {\BGO} and 7.40 eV \cite{franchy1997growth} for {\TAO}.
An ordered AlGaO$_3$ alloy is also investigated, with all Al atoms on octahedral sites and all Ga atoms on tetrahedral sites \cite{peelaers2018structural,*peelaers2019erratum}.
{\color{black} The use of an ordered structure serves as an approximation; the computation of structures representative of disordered alloys would be computationally prohibitive.  We note that for the 50\% Al content, ordered AlGaO$_3$ was found to be a line compound due to the preference for Al to occupy octahedral sites \cite{peelaers2018structural,*peelaers2019erratum}.}
Spin polarization is included and full structural optimizations were performed.
The optimized computed structural parameters of {\BGO}, AlGaO$_3$ and {\TAO} are summarized in Table~\ref{tab:structure}; they compare favorably with experiment.

\begin{table}
\caption{Structural parameteres (lattice parameters, \AA; angle $\beta$, degrees), formation enthalpy per formula unit ($\Delta H_f$, eV/f.u.),  and band gaps ($E_\text{gap}$, eV) for monoclinic {\GO}, AlGaO$_3$ and {\AO}.
Experimental  results are also listed for comparison.}
\begin{tabular}{ccccccc}
\hline
    & \multicolumn{2}{c}{\GO}& \multicolumn{2}{c}{AlGaO$_3$} & \multicolumn{2}{c}{\AO} \\
\hline
    &Calc & Expt & Calc & Expt & Calc & Expt \\
\hline
$a$ & 12.14 & 12.21$^a$ &11.86 & 12.00$^a$ & 11.66 & 11.85$^e$   \\
$b$ & 3.02 & 3.04$^a$ & 2.94 & 2.98$^a$ & 2.88 & 2.90$^e$ \\
$c$ & 5.78 & 5.81$^a$ & 5.69 & 5.73$^a$ & 5.57 & 5.62$^e$  \\
$\beta$ & 103.77 & 103.87$^a$ &104.25  & 104.03$^a$& 104.04 & 103.83$^e$ \\
\hline
$\Delta H_f$  &$-$10.22  & $-$11.29$^b$ & $-$13.87 & & $-$16.11 &  \\
\hline
$E_\text{gap}^\text{indir}$ & 4.83 &  &  5.81 &  & 7.41 & 7.40$^f$  \\
$E_\text{gap}^\text{dir}$ & 4.87 & 4.76$^c$, 4.88$^d$ &  5.89 & & 7.74 & \\
\hline
\end{tabular}
\\
{$^a$Ref.~\onlinecite{kranert2015lattice};}
{$^b$Ref.~\onlinecite{lide2004crc};}
{$^c$Ref.~\onlinecite{matsumoto1974absorption};}
{$^d$Ref.~\onlinecite{sturm2016dipole};}
{$^e$Ref.~\onlinecite{zhou1991structures};}
{$^f$Ref.~\onlinecite{franchy1997growth};}
\label{tab:structure}
\end{table}

A 120-atom supercell was constructed for the point-defect calculations, as shown in Fig.~\ref{fig1:struct}(b).
Selected tests were performed using 160-atom supercells.
{\color{black}Tests for larger supercells are too expensive to perform with the hybrid functional, but supercell size convergence was further checked using the generalized gradient approximation of Perdew, Burke, and Ernzerhof (PBE) \cite{Perdew1996} for Si and C impurities in supercells containing up to 1280 atoms.
Transition levels were found to change by less than 0.1 eV.}
For each point defect, we perform multiple calculations with different {\color{black} symmetry-breaking} initial local distortions around the defect in the initial state, and optimize the atomic positions until the Hellmann-Feynman forces are lower than 5 meV/\AA.
{\color{black}All the supercell calculations are consistently performed at the lattice parameters obtained with the same plane-wave energy cutoff, as listed in Table~\ref{tab:structure}.}

Migration barriers are calculated using the climbing-image nudged elastic band (cNEB) method \cite{henkelman2000climbing}.
To mitigate computational cost we perform one-shot HSE calculations for the migration barriers ($E_\text{b}$): we use PBE \cite{Perdew1996} in the cNEB calculations, followed by static HSE total energy calculations based on the initial and barrier geometries.
We tested the accuracy of this approach for the migration of H$_i^+$ in {\AO} and {\GO} along the [010] direction; full HSE calculations yield migration barriers that are within 0.1 eV of the results obtained using one-shot HSE calculations.

After determining the migration barrier of a defect, we can estimate the temperature at which the defect becomes mobile with transition state theory \cite{vineyard1957frequency}. This temperature is denoted as an ``annealing temperature'', above which the defect is in thermodynamical equilibrium.  In transition state theory \cite{vineyard1957frequency}, the rate $\Gamma$ at which the defect hops to a neighboring site can be expressed as
\begin{equation}\label{anneal}
\Gamma = \Gamma_0 \exp (-\frac{E_\text{b}}{k_BT}) \, ,
\end{equation}
where $k_B$ is the Boltzmann constant and $E_\text{b}$ the migration barrier. The prefactor $\Gamma_0$ is an effective frequency associated with the vibration of the defect; for the H interstitial in both {\GO} and {\AO}, we choose $\Gamma_0=10^{14}$ $s^{-1}$, which is approximately the dominant O-H vibrational frequency {\GO} \cite{weiser2018structure,varley2011hydrogenated}.   We estimate the annealing temperature as the temperature at which the rate $\Gamma = 1$ s$^{-1}$ \cite{janotti2007native}.
We note that the annealing temperature is not very sensitive to the choice of $\Gamma_0$.

\subsection{Formation energy}\label{formE}
The formation energy of a defect is used to assess the likelihood of the presence of the defect and its concentration.
For Si substituting on an Al site (Si$_\text{Al}$), the formation energy is calculated as
\begin{equation}
\begin{split}
        E^f(\text{Si}^q_\text{Al}) & = E_\text{tot}(\text{Si}^q_\text{Al})-E_\text{tot}(\text{Al}_2\text{O}_3) -(\mu_\text{Si}+\mu^0_\text{Si}) \\
        & + (\mu_\text{Al}+\mu^0_\text{Al})+q(E_\text{F}+E_\text{VBM}) + \Delta^q ,
\end{split}
\end{equation}
where $E_\text{tot}(\text{Si}^q_\text{Al})$ is the total energy of one Si$_\text{Al}$ in charge state $q$ in the supercell, $E_\text{tot}(\text{Al}_2\text{O}_3)$ is the total energy of the bulk supercell, and $E_\text{F}$ is the Fermi energy, referenced to the valence-band maximum (VBM).
$\Delta^q$  is a finite-size correction term for charged defects~\cite{freysoldt2009fully, freysoldt2011electrostatic}.
{\color{black}We adopt the previously reported values for the dielectric constants of {\GO}, AlGaO$_3$, and {\AO}~\cite{varley2020prospects}}. 
The chemical potentials are referenced to the elemental phases, e.g., $\mu^0_\text{Si}=E_\text{tot}(\text{Si})$ and $\mu^0_\text{Al}=E_\text{tot}(\text{Al})$.

The Al and O chemical potentials have to fulfill the stability condition for bulk {\AO}:
\begin{equation}
    2\mu_\text{Al}+3\mu_\text{O}= \Delta H_f(\text{Al}_2\text{O}_3) ,
\end{equation}
where $\Delta H_f(\text{Al}_2\text{O}_3$) is the formation enthalpy of bulk {\TAO}, as shown in Table ~\ref{tab:structure}.
The calculated formation enthalpies of {\BGO} and AlGaO$_3$ are listed in Table ~\ref{tab:structure} as well.
$\mu_\text{Al} = 0$ corresponds to Al-rich conditions, and $\mu_\text{O} = 0$ to O-rich (Al-poor) conditions.
For purposes of presenting results for the impurities, we choose the chemical potentials to correspond to the solubility limit, i.e., the highest value of the chemical potential that is compatible with formation of other phases that can result from interactions of the impurity with the host elements.
For $\mu_\text{Si}$, this correspond to equilibrium with SiO$_2$:
\begin{equation}
    \mu_\text{Si}+2\mu_\text{O} = \Delta H_f(\text{Si}\text{O}_2) \, ,
\end{equation}
where $\Delta H_f(\text{Si}\text{O}_2)$ is the calculated formation enthalpy of SiO$_2$.
For C and H, the upper limits depend on the host chemical potentials:
the limits for $\mu_\text{C}$ correspond to Al$_4$CO$_4$ for Al-rich and CO$_2$ for Al-poor conditions;
the limits for $\mu_\text{H}$ correspond to H$_2$ for Al-rich and H$_2$O for Al-poor.

To be able to comment on the behavior in {\ALGO} alloys, we also calculate formation energies in {\GO}.
The limiting phases for Si and H are the same as those in the {\AO} case.
The limits for $\mu_\text{C}$ in {\GO} correspond to graphite for Ga-rich and CO$_2$ for Ga-poor conditions.
We note that this is different from Ref.~\onlinecite{lyons2014carbon}, where diamond was assumed as the limiting phase under Ga-poor conditions.

The charge-state transition level between charge states $q$ and $q'$, denoted as ($q/q'$), is calculated based on the formation energies:
\begin{equation}\label{eq_levels}
    (q/q') =\frac{E^f(\text{Si}^q_\text{Al}; E_\text{F}=0)-E^f(\text{Si}^{q'}_\text{Al}; E_\text{F}=0)}{(q'-q)} \, .
\end{equation}

\subsection{Stability of \textit{DX} centers}
\label{sec:stab}

Following Chadi and Chang \cite{chadi1988theory}, we label the neutral and  positively charged donors on the substitutional
site $d^0$ and $d^{+}$, respectively. $DX^{-}$ is used to denote the configuration in the negative charge state.
The formation of a $DX$ center leads to self-compensation, as described by the process
\begin{equation}
2d^0 \longrightarrow d^+ + DX^{-} \, .
\label{eq:d}
\end{equation}
The \textit{DX} center can be characterized by the effective correlation parameter $U$, defined by
\begin{equation}\label{eq_u}
    U = E^f(d^+)+E^f(DX^{-}) -2 E^f(d^0) \, .
\end{equation}
Stability of the \textit{DX} center is typically characterized by a negative value of $U$.
However, using $U$ as the descriptor of the \textit{DX} center can be problematic because of the challenges involved in accurately calculating the energy of the neutral charge state when the (+/0) level (and hence the Kohn-Sham state for $d^0$) is near or above the conduction-band minimum (CBM).
Instead, we will use the ($+/-$) charge-state transition level as a descriptor: our criterion will be that the \textit{DX} center is stable if the ($+/-$) level lies below the CBM.

\section{Results and Discussion} \label{mainresult}

\begin{figure}
\includegraphics[width=0.52\textwidth]{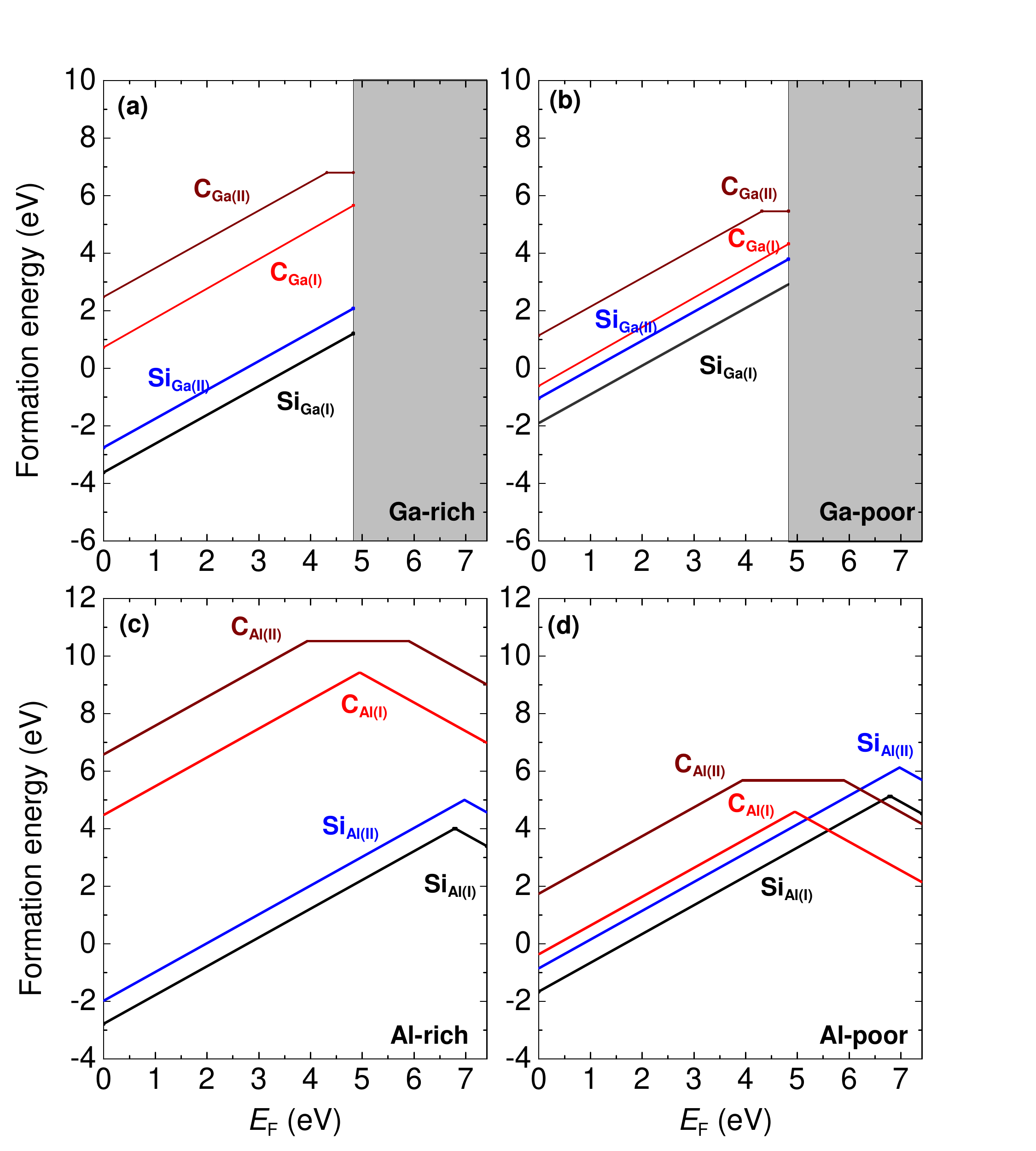}
\caption{\label{fig:formation}
Formation energy versus Fermi level for Si and C impurities in (a)-(b) {\BGO} and (c)-(d) {\TAO}. (a) and (c) are for cation-rich, and (b)-(d) for cation-poor conditions.
The grey area indicates the conduction band of {\BGO}.
}
\end{figure}

\subsection{Silicon} \label{siresult}

The formation energies of Si on the tetrahedral (Si$_\text{Al(I)}$) and octahedral (Si$_\text{Al(II)}$) sites in {\TAO} are shown in Fig.~\ref{fig:formation}(c,d) for cation-rich and cation-poor conditions. For completeness and consistency, we also show the formation energies of Si in {\GO} in Fig.~\ref{fig:formation}(a,b); these have been calculated previously~\cite{varley2010oxygen} and  our results are consistent with the previous results.
Note that in Fig.~\ref{fig:formation} (and all subsequent formation-energy figures) we keep the horizontal-axis scale the same for {\BGO} and {\TAO}, which facilitates comparisons of formation energies.
In principle, we could also take the valence-band alignment between {\BGO} and {\TAO} into account to more accurately compare charge-state transition levels~\cite{lyons2014cnitrides,lyons2017npjgan}; however, since the valence-band offset is relatively small~\cite{peelaers2018structural,*peelaers2019erratum}, we do not include this alignment.

Similar to {\BGO}, Si prefers to be incorporated on the tetrahedral site in {\TAO}: $E^f$(Si$^+_\text{Al(I)}$) is lower than $E^f$(Si$^+_\text{Al(II)}$) by 0.80 eV.
In {\BGO} Si behaves as a shallow donor on either site, i.e., only the positive charge state is stable for all Fermi-level positions within the band gap.
In {\TAO} a ($+/-$) charge-state transition level occurs in the band gap, at 0.62 eV below the CBM for Si$_\text{Al(I)}$ and at 0.43 eV below the CBM for Si$_\text{Al(II)}$ (see Table~\ref{tab_sum}).
Table~\ref{tab_sum} also lists values for the other transition levels.
For Si$_\text{Al(I)}$, there is a small region of stability for the neutral charge state; we found ($+/0$) and ($0/-$) transition levels at 6.77 eV and 6.81 eV, respectively, corresponding to $U$=0.04 eV.
For Si$_\text{Al(II)}$, the ($+/0$) level lies above the ($0/-$) level and we have a negative-$U$ center (i.e., the neutral charge state is never thermodynamically stable), with $U$=$-$0.44 eV.

\begin{table}
\caption{Charge-state transition levels (eV) [Eq.~(\ref{eq_levels})] and effective correlation parameter $U$ (eV) [Eq.~(\ref{eq_u})] for various impurities in {\BGO}, ordered AlGaO$_3$, and {\TAO}.
The neutral and negative charge states that are used to compute charge-state transition levels and $U$ all correspond to localized states.
Values for Si in {\GO} are missing since the localized neutral and negative charge states cannot be stabilized.
We also list $x^\text{onset}$ (\%), the Al concentration in {\ALGO} corresponding to the onset of \textit{DX} behavior. }
\begin{ruledtabular}
\begin{tabular}{cccccc}
{\BGO} &  ($+/0$)  & ($+/-$) 	& ($0/-$) 	& $U$ 	&	$x^\text{onset}$ 	 \\
\hline
C$_\text{Ga(I)}$ & 5.20	& 4.92 &	4.65 &	$-$0.54   \\
C$_\text{Ga(II)}$ & 4.32	& 5.05 &	5.77 &	1.46   \\
H$_\text{i}$ & 5.47 & 	4.84 & 	4.21 & 	$-$1.25   \\
H$_\text{O(I)}$ & 5.47	 & 5.20 & 	4.91 &	$-$0.55   \\
H$_\text{O(II)}$ & 4.98	& 5.01	& 5.04 & 	0.07  \\
H$_\text{O(III)}$ & 5.71	& 5.56	& 5.42	& $-$0.29   \\
\hline
AlGaO$_3$ &  ($+/0$)  & ($+/-$) 	& ($0/-$) 	& $U$ 	&	 	 \\
\hline
Si$_\text{Ga(I)}$ & 6.39	& 6.19 &	6.00 &	$-$0.39  &  \\
Si$_\text{Al(II)}$ & 6.065	&6.46 &	6.85 &	0.79  & \\
C$_\text{Ga(I)}$ & 4.93 &	4.81 &	4.69	& $-$0.24  &  \\
C$_\text{Al(II)}$ & 4.05 &	4.97 &	5.90	& 1.86   & \\
H$_\text{i}$ & 5.70 &	4.67	& 3.65	& $-$2.05  &  \\
H$_\text{O(I)}$ & 5.92	& 5.69 & 	5.45 &	$-$0.47 &   \\
H$_\text{O(II)}$ & 5.60 &	5.21 &	4.83 &	-0.77 &   \\
H$_\text{O(III)}$ & 5.69 &	5.70 &	5.71 &	0.02  & \\
\hline
{\TAO} &  ($+/0$)  & ($+/-$) 	& ($0/-$) 	& $U$ 	&	 	 \\
\hline
Si$_\text{Al(I)}$ & 6.77	& 6.79  &	6.81 &	0.04 & 70\% \\
Si$_\text{Al(II)}$ & 7.20 & 	6.98 &	6.76	& $-$0.44 & 81\% \\
C$_\text{Al(I)}$ & 4.96	& 4.95 &	4.95	& $-$0.01 & 5\%  \\
C$_\text{Al(II)}$ & 4.23	& 5.07 &	5.90	& 1.95 & 51\%  \\
H$_\text{i}$ & 5.83 &	4.83 &	3.84	& $-$1.99 & 1\%  \\
H$_\text{O(I)}$ &6.51 &	6.36 &	6.41	&$-$0.10 & 37\%  \\
H$_\text{O(II)}$ & 5.78&	5.70 &	 5.61&	$-$0.16 & 13\%	  \\
H$_\text{O(III)}$ & 6.11 &	6.12	& 6.13 &	0.03 & 41\%  \\
\end{tabular}
\end{ruledtabular}
\label{tab_sum}
\end{table}

\begin{figure*}
\includegraphics[width=0.95\textwidth]{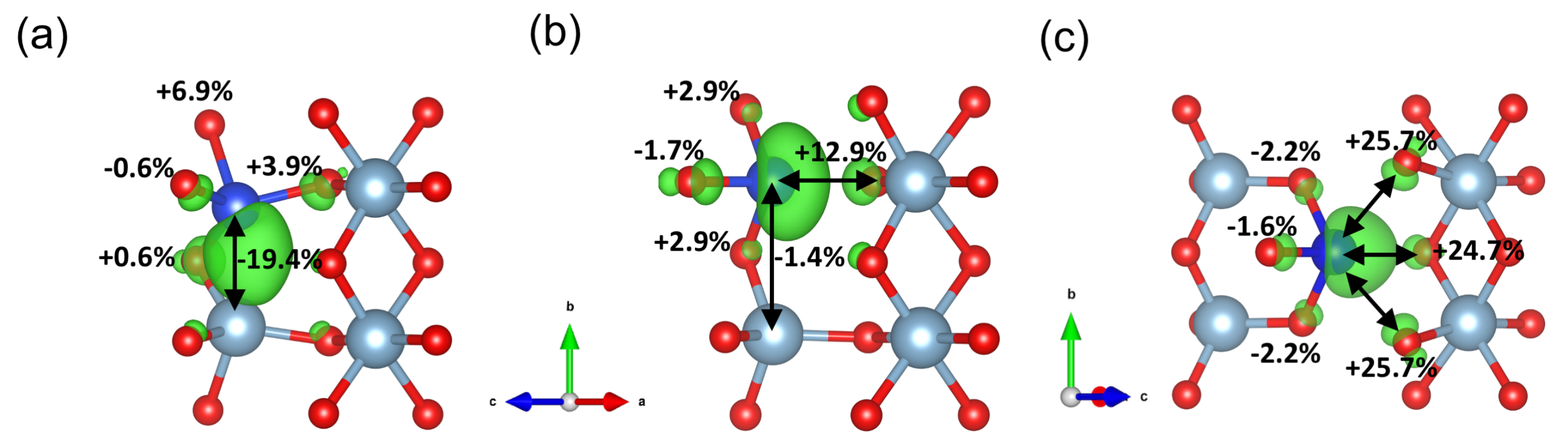}
\caption{\label{fig:si_local}
Local structure and localized charge density for negatively charged \textit{DX} configurations of (a) ground-state Si$^-_\text{Al(I)}$, (b) metastable-state Si$^-_\text{Al(I)}$, and (c) Si$^-_\text{Al(II)}$ in {\TAO}.
Dark blue spheres denote Si, light blue Al, and red O.
Percentage changes of bond lengths, referenced to bonds in bulk {\AO}, are indicated.
Isosurfaces of the charge density (0.02 $e$/\AA$^3$) of the localized state in the \textit{DX} centers are shown in green.}
\end{figure*}

For Fermi levels in the upper part of the gap, Si$^-_\text{Al}$ is in a negative charge state, associated with the formation of a \textit{DX} center.
For Si$_\text{Al(I)}$ we found that \textit{two} locally stable \textit{DX} configurations exist.
The local geometry and the charge density of the localized electrons are illustrated in Fig.~\ref{fig:si_local}.
Figure~\ref{fig:si_local}(a) corresponds to the ground state of Si$^-_\text{Al(I)}$, while Fig.~\ref{fig:si_local}(b) represents a metastable state that is 0.40 eV higher in energy.
The metastable state [Fig. \ref{fig:si_local}(b)] features the type of broken bond that is often considered characteristic of a \textit{DX} center, resulting in a dangling bond on the Si atom that is filled with two electrons.
The ground state of Si$^-_\text{Al(I)}$ [Fig. \ref{fig:si_local}(a)] also displays significant bond distortions, both in the SiO$_4$ tetrahedron and its nearest AlO$_4$; however, no clear bond breaking occurs.
Instead, the main effect of the distortion is to reduce the Si-Al distance from 2.88 \AA\ (the distance in bulk {\TAO}) to 2.32 \AA, creating a two-electron bond between Si and Al, as clearly seen in Fig.~\ref{fig:si_local}(a).
Octahedrally coordinated Si$^-_\text{Al(II)}$ also forms a \textit{DX} center, shown in Fig.~\ref{fig:si_local}(c). Three Si-O bonds are broken in the SiO$_6$ octahedron, and two electrons are localized on the Si dangling bond.
Tests for the Si \textit{DX} center in a 160-atom supercell yielded very similar local geometries, with bond lengths differing by less than 0.01 {\AA}.
The 160-atom supercell calculations yield a ($+/-$) charge-state transition level that is within 0.05 eV of the result using a 120-atom supercell.

The formation of the silicon \textit{DX} center indicates that Si cannot be a shallow donor in {\TAO}.
To gain insight into its effectiveness in {\ALGO} alloys, we evaluate the position of the ($+/-$) transition level and compare it with the CBM in {\ALGO} alloys.
The negative charge state employed to compute the ($+/-$) transition level has to correspond to a localized state.
Such a localized state can be stabilized in {\AO} and ordered AlGaO$_3$ alloy but not in {\GO}.
Therefore, the ($+/-$) transition levels in alloys are interpolated (or extrapolated towards {\GO}) based on the values in {\AO} and ordered AlGaO$_3$ alloy (see values in Table~\ref{tab_sum}).
The values of the CBM are interpolated based on those in {\GO} and {\AO}, with the addition of bowing using a bowing parameter of 0.93 eV and assuming all of the band-gap bowing occurs in the CBM~\cite{peelaers2018structural,*peelaers2019erratum}.

\begin{figure}
\includegraphics[width=0.52\textwidth]{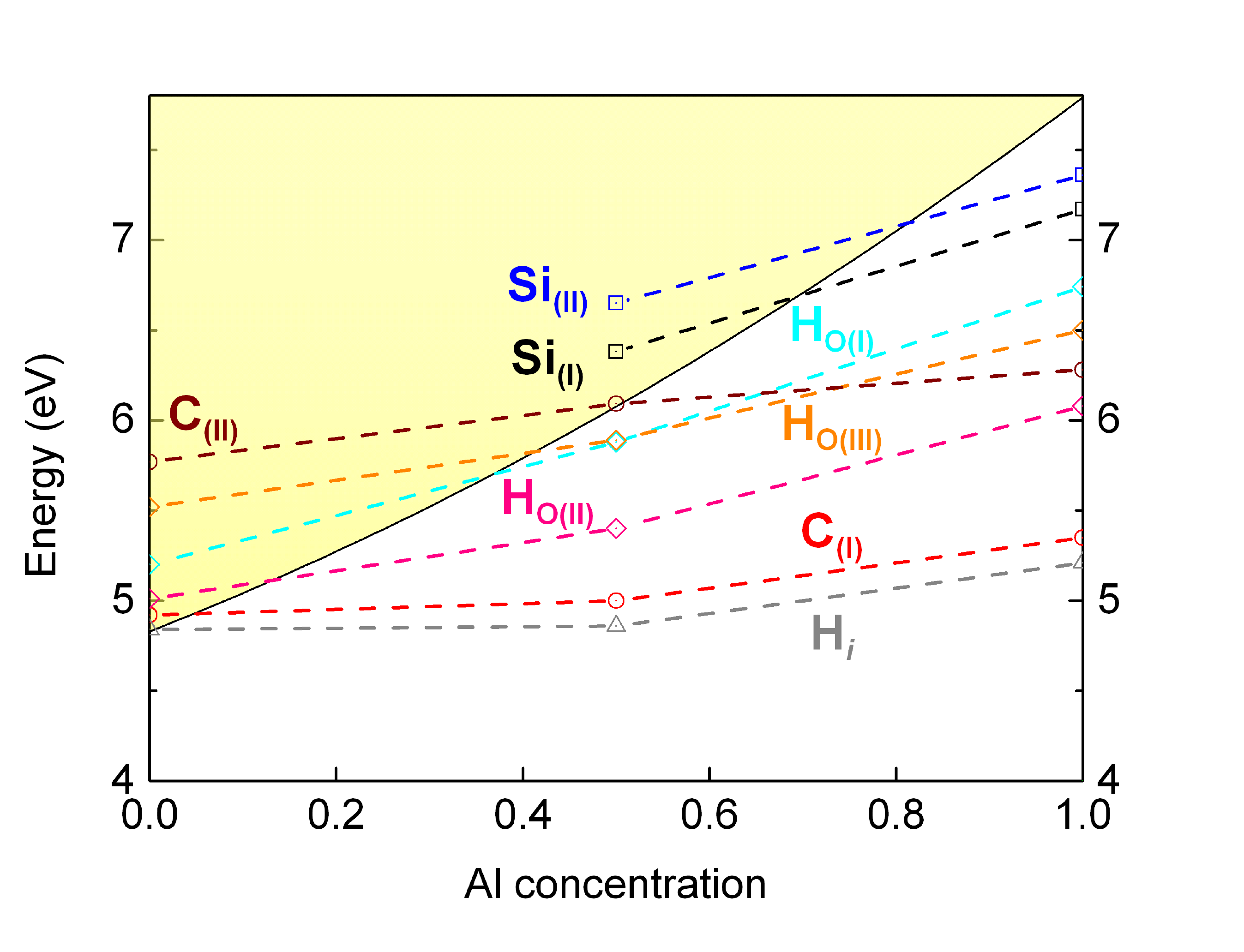}
\caption{\label{fig:band}
The ($+/-$) levels of Si$_\text{(I)}$, Si$_\text{(II)}$, C$_\text{(I)}$, H$_i$, H$_\text{O(I)}$, H$_\text{O(II)}$, H$_\text{O(III)}$, and the ($0/-$) level of C$_\text{(II)}$ obtained from calculations in {\AO}, {\GO} and ordered AlGaO$_3$ alloys.
The notation Si$_\text{(I)}$, resp. Si$_\text{(II)}$, denotes substitutional Si on the tetrahedral (I), resp. octahedral (II) cation site, and similarly for C$_\text{(I)}$ and C$_\text{(II)}$.
The charge-state transition levels for Si$_\text{I}$ and Si$_\text{II}$ in {\GO} are missing since the localized neutral and negative charge states cannot be stabilized.
The dashed lines connecting the calculated values are just a guide to the eye.
The CBM/VBM band offsets as a function of the Al concentration for alloys were obtained from Ref.~\protect\onlinecite{peelaers2018structural,*peelaers2019erratum}.
}
\end{figure}

The resulting values of $x^\text{onset}$, the Al concentration at which the ($+/-$) level moves below the CBM (i.e., the \textit{DX} center becomes stable) are listed in Table~\ref{tab_sum}.
We found Si incorporating on the tetrahedral site will be an effective donor over a wide range of Al alloying, up to 70\% Al (see Fig.~\ref{fig:band}).
This value is somewhat lower than the $x^\text{onset}$=86\% reported in Ref.~\onlinecite{varley2020prospects}.
The difference can be attributed to the fact that our calculated ($+/-$) level in {\AO} (AlGaO$_3$) is 0.12 eV (0.37 eV)  lower than that in Ref.~\onlinecite{varley2020prospects}, due to the fact that we identified a lower-energy structure for the \textit{DX} center.
In addition, we find that Si on the octahedral site is also an effective donor in {\ALGO}, up to 81\% Al (see Fig.~\ref{fig:band}).

{\color{black}Our treatment of defect behavior in the alloy by performing linear interpolation (Fig.~\ref{fig:band}) is obviously an approximation.
In the case of {\ALGO} alloys, it is likely to be quite reliable, due to the similarity in the properties of Al and Ga cations;
indeed, we have found that the geometries of defect configurations are very similar in {\BGO} and {\TAO}.
Some degree of validation is also obtained from the results for transition levels explicitly calculated for AlGaO$_3$ alloys.
As shown in Fig.~\ref{fig:band}, the results for the alloy generally fall on the line that linearly interpolates between {\BGO} and {\TAO};
the small deviations that occur for C$_{\rm (I)}$ and H$_i$ impurities amount to less than 0.15 eV, which is gratifying agreement.
We also note that this interpolation procedure has been demonstrated to be effective for obtaining transition levels in AlGaN alloys~\cite{mccluskey1998metastability,gordon2014hybrid}. Indeed, experiment ~\cite{skierbiszewski1999evidence} indicated that the O$_\text{N}$ transition level in AlGaN alloys varies \textit{linearly} with alloy concentration, supporting the validity of linear interpolation.  
}

\subsection{Carbon} \label{cresult}

\subsubsection{Incorporation}
\label{sec:cinc}

Carbon is a commonly observed background impurity in semiconductors, particularly in MOCVD-grown films.
In {\GO}, carbon on the tetrahedral site is a shallow donor \cite{lyons2014carbon,bouzid2019defect}.
Here we investigate whether the increase in band gap in {\ALGO} will cause the carbon ($+/-$) level to drop below the CBM, turning carbon into a compensating acceptor.
We also study whether carbon on the oxygen site could cause compensation.

Lyons \textit{et al.}~\cite{lyons2014carbon} investigated carbon impurities on the anion, cation, and interstitial sites in {\GO} and showed that under most conditions, the carbon impurity prefers the substitutional tetrahedral cation site.
For a consistent comparison, we reproduced the formation energies of carbon impurities on the cation sites in {\GO}[see Figs.~\ref{fig:formation}(a,b)] and our results agree with Lyons \textit{et al.}~\cite{lyons2014carbon} under the same condition.
Similar to the case of {\GO}, C prefers to occupy the tetrahedral site (C$^+_\text{Al(I)}$) in monoclinic {\AO}, with a formation energy 2.10 eV lower than that of C$^+_\text{Al(II)}$ at the VBM (see Fig.~\ref{fig:formation}).
At the CBM, the formation energy of C$^-_\text{Al(I)}$ is 2.03 eV lower than that of C$^-_\text{Al(II)}$.
The behavior of C$^-_\text{Al}$ as a compensating center is broadly consistent with previous results for carbon in $\alpha$-{\AO}, the corundum phase~\cite{choi2013impact}.
Figure~\ref{fig:formation} also shows that under Al-rich conditions the formation energy of C$^-_\text{Al(I)}$ is high, significantly decreasing when we move to Al-poor conditions.
However, carbon incorporation in MOCVD-grown samples is likely not governed by thermodynamic equilibrium, but rather determined by unintentional incorporation due to incomplete dissociation of metal-organic precursors.
In the following we therefore examine both C$_\text{Al(I)}$ and C$_\text{Al(II)}$.

\subsubsection{Carbon on the tetrahedral site}
\label{sec:ctet}

\begin{figure*}
\includegraphics[width=0.95\textwidth]{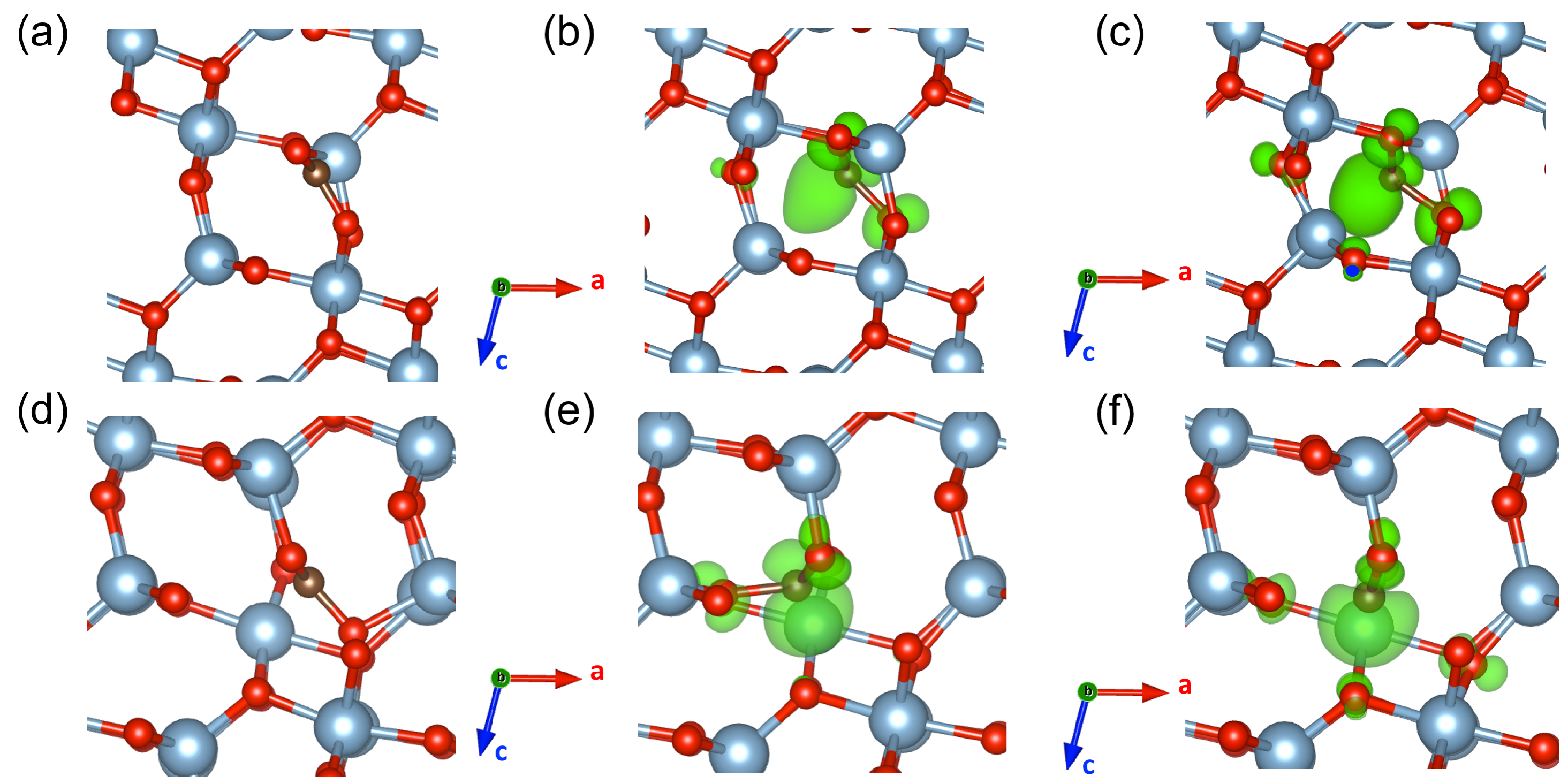}
\caption{\label{fig:c_local}
Local structure and localized charge density for (a) metastable 3-fold-coordinated configuration of C$^+_\text{Al(I)}$, (b) C$^0_\text{Al(I)}$, (c) C$^-_\text{Al(I)}$, (d) C$^+_\text{Al(II)}$, (e) C$^0_\text{Al(II)}$ and (f) C$^-_\text{Al(II)}$ in {\TAO}.
Brown spheres denote C, light blue Al, and red O.
Isosurfaces of the charge density (0.01 $e/\text{\AA}^3$) of the localized electrons in C$^0_\text{Al(I)}$, C$^-_\text{Al(I)}$, C$^0_\text{Al(II)}$ and C$^-_\text{Al(II)}$ are shown in green.}
\end{figure*}

The small atomic size of carbon leads to bond breaking and a variety of possible local geometries, as depicted in Fig.~\ref{fig:c_local}.
Focusing first on C$_\text{Al(I)}$, already in the positive charge state we find two competing configurations of C$^+_\text{Al(I)}$.
In the most stable state (not depicted), C forms four C-O bonds, and
the oxygen cage shrinks, yielding four C-O bonds with bond lengths ranging from 1.40 to 1.44 {\AA}.
In the metastable configuration [see Fig.~\ref{fig:c_local}(a)], which is only 87 meV higher than the ground state, the C-O(III) bond is broken, and the
remaining three C-O bonds have similar lengths (1.28, 1.30, 1.30 {\AA}) and almost lie in the same plane, with nearly 120$^\circ$ angles between the bonds.
Carbon effectively forms an $sp^2$-bonded configuration with its three O neighbors.
This 3-fold coordination was also found for C$_\text{Ga(I)}$ in {\GO}~\cite{lyons2014carbon}, with very similar C-O bond lengths (1.27--1.30 {\AA}).
The 3-fold-coordinated configuration is the lowest-energy configuration in {\GO} and AlGaO$_3$.

For the neutral charge state of C$_\text{Al(I)}$, we found that we can stabilize as many as five distinct atomic configurations, starting from initial structures in which different C-O bonds are broken in a CO$_4$ tetrahedron.
The most stable configuration corresponds to breaking of the C-O(III) bond, forming a nonplanar CO$_3$ cluster, as shown in Fig.~\ref{fig:c_local}(b).
The C-O bond lengths are 1.35, 1.39, 1.39 \AA, while the distance between C and O(III) is 2.82 \AA.
This configuration is similar to the 3-fold coordination for the positive charge state, but the carbon has moved beyond the plane of its oxygen neighbors.
As seen in Fig.~\ref{fig:c_local}(b), the C has moved towards a next-nearest Al(I) neighbor in the same (010) plane, reducing the C-Al distance from 3.5 {\AA} (for C on the nominal Al site) to 2.6 {\AA}.
In the process, the charge of the localized state becomes confined between C and that next-nearest-neighbor Al atom.
The fact that this state is localized clearly indicates that C$_\text{Al(I)}$ is not a shallow donor in {\TAO}.
Indeed, the corresponding Kohn-Sham state lies 4.42 eV below the CBM.

Another configuration in which the same C-O(III) bond is broken can be stabilized with a distance of 2.38 \AA\ between C and O(III); this configuration is only 0.05 eV higher in energy.
Other configurations involve breaking a C-O(I) bond (0.80 eV higher than the ground state) or  breaking a C-O(II) bond (0.91 eV higher than the ground state).
The geometry in which all four C-O bonds in the CO$_4$ tetrahedron are maintained is also a local minimum, 0.89 eV higher in energy than the ground-state C-O(III) bond-breaking configuration.

In the negative charge state, the most stable state is similar to the ground state of C$^0_\text{Al(I)}$, with a broken C-O(III) bond [see Fig.~\ref{fig:c_local}(c)].
Other configurations can be stabilized, but with energies that are at least 1.3 eV higher than the ground state.
Figure~\ref{fig:c_local}(c) depicts the charge density of the localized state.
Comparison with Fig.~\ref{fig:c_local}(b) indicates that with the addition of an extra electron a C-Al bond has formed with the next-nearest-neighbor Al(I).
The C-Al distance is reduced to 2.0 \AA\ after relaxation.
This is another example of a $DX$ configuration in which a cation-cation bond forms, but we note the distinction with the \textit{DX} configuration of Si$^-_\text{Al(I)}$ shown in Fig.~\ref{fig:si_local}(a):
Si$^-_\text{Al(I)}$ bonds with a \textit{nearest-neighbor} tetrahedral host cation, while
C$^-_\text{Al(I)}$ bonds with a \textit{next-nearest-neighbor} tetrahedral host cation.
A similar bonding configuration for the negative charge state of carbon on the tetrahedral site is also observed in {\GO} and AlGaO$_3$.

\subsubsection{Carbon on the octahedral site}
\label{sec:coct}

Turning now to C$_\text{Al(II)}$,  a variety of competing local geometries with different broken bonds are also present in the positive, neutral and negative charge states due to the small atomic size of carbon. In the positive charge state, the ground state corresponds to bond breaking of two C-O(III) bonds and one C-O(I) bond in the nominal octahedron, leaving the remaining three C-O bonds  (1.28, 1.28, 1.30 {\AA}) to form a 3-fold coordinated configuration that is similar to the metastable configuration of C$^+_\text{Al(I)}$ [see Fig.~\ref{fig:c_local}(d)]. We also observe a competing metastable state of C$^+_\text{Al(II)}$ with a different 3-fold coordinated configuration that exhibits bond breaking of all three C-O(III) within the octahedron. This metastable state is 0.20 eV higher than the ground state.
This 3-fold coordinated configuration with three broken C-O(III) bonds becomes the ground state in the neutral charge state [see Fig.~\ref{fig:c_local}(e)].
The remaining three C-O bonds form a nonplanar CO$_3$ cluster, with C-O bond lengths of 1.37, 1.37, 1.41 \AA. The charge of the localized state, dominated by a $p$-orbital character, is confined between C and the center of the octahedron.
Another (metastable) 3-fold coordinated configuration with bond breaking of two C-O(III) and one C-O(II) is 0.28 eV higher.

In the negative charge state of C$_\text{Al(II)}$, the most stable configuration is 2-fold coordinated, with two equivalent C-O(I) bonds (1.40, 1.40 \AA) [see Fig.~\ref{fig:c_local}(f)]. The localized states exhibit roughly equal $p$- and $s$-orbital character and the associated charges are confined near C, with an appreciable weight at the center of the octahedron.  We also find another competing 2-fold coordinated metastable configuration, with one C-O(I) bond (1.26 \AA) and one C-O(II) bond (1.91 \AA) remaining, and with an energy 0.18 eV higher.

\subsubsection{Carbon on cation sites as a compensating center}
\label{sec:ccom}

When discussing substitutional C in {\ALGO} alloys, C could be replacing either a Ga or an Al atom; in this section (as in Fig.~\ref{fig:band}) we use the generic notation C$_\text{(I)}$ to denote either C$_\text{Ga(I)}$ or C$_\text{Al(I)}$, and C$_\text{(II)}$ to denote either C$_\text{Ga(II)}$ or C$_\text{Al(II)}$.
C$_\text{(I)}$ is a negative-$U$ center across {\GO}, {\AO} and the ordered  {\ALGO} alloy (see Table~\ref{tab_sum}).  We can evaluate its ($+/-$) transition level in the relevant range of {\ALGO} alloy compositions by interpolating between {\GO} and the AlGaO$_3$ alloy, with consideration of the band bowing \cite{peelaers2018structural,*peelaers2019erratum}  (see Fig.~\ref{fig:band}).
For C$_\text{(I)}$, we find that the impurity starts acting as an acceptor in {\ALGO} alloys already at 5\% Al.
C$_\text{(II)}$ is \textit{not} a negative-$U$ center, but it can still act as compensating acceptor in $n$-type {\AO} (see Fig.~\ref{fig:formation}). To estimate the onset of C$_\text{(II)}$ as a compensator in {\ALGO} alloys, we interpolate the ($0/-$) transition levels between {\AO} and the AlGaO$_3$ alloy  (see Fig.~\ref{fig:band}), finding that C$_\text{(II)}$ starts acting as an acceptor at 51\% Al.

Our results indicate that unintentional C incorporation is a potential explanation for the lack of doping efficiency observed at low concentrations of Si in MOCVD-grown {\ALGO} alloys.
Since unintentionally incorporated C$_\text{(I)}$ acts as a compensating acceptor in {\ALGO} with $x\geq 5\%$, the Si concentration needs to exceed the concentration of C$_\text{(I)}$ in order for Si to actually provide $n$-type doping.

\subsubsection{Carbon on oxygen sites, $\mathrm{C}_\mathrm{O}$}
\label{sec:co}

\begin{figure}
\includegraphics[width=0.52\textwidth]{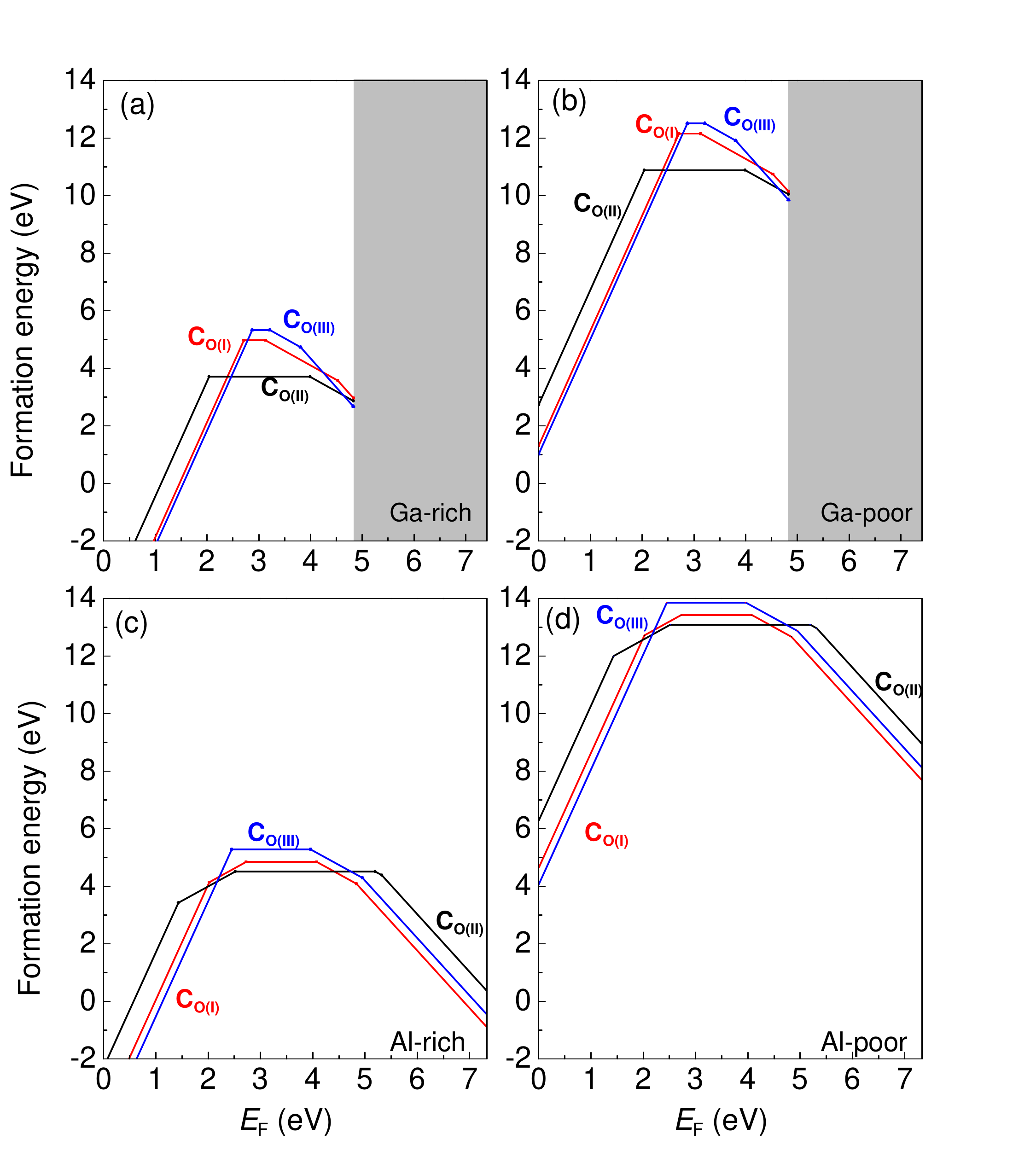}
\caption{\label{fig:co}
Formation energy versus Fermi level for C$_\mathrm{O}$ in (a)-(b) {\BGO} and (c)-(d) {\TAO}. (a) and (c) are for cation-rich, and (b)-(d) for cation-poor conditions.  Carbon incorporation on three possible O sites is considered: O(I), O(II), and O(III).
The grey area indicates the conduction band of {\BGO}.
}
\end{figure}

In {\GO}, it was found that C$_\mathrm{O}$ acts as a compensating acceptor
when the Fermi level is high in the gap \cite{lyons2014carbon}.
Reference~\onlinecite{lyons2014carbon} reported results for a single O site and only for O-rich conditions; in Figs.~\ref{fig:co}(a,b) we report a full set of results for C$_\mathrm{O}$ in {\GO}.
{\color{black}Under conditions relevant for $n$-type doping, the formation energy of C$_\mathrm{O}$ is lowest when the Fermi level is at the CBM under Ga-rich conditions}; however, even under these extreme conditions, this formation energy is still 2.69 eV.

The formation energy of C$_\mathrm{O}$ can be significantly lower in {\AO}, as shown in Figs.~\ref{fig:co}(c,d).
For Fermi levels low in the gap (which may be difficult to attain in {\AO}), C$_\mathrm{O}$ prefers the 4+ charge state.
At higher Fermi-level positions, the neutral, 1$-$, and 2$-$ charge states become more favorable.
For Fermi levels close to the CBM, C$_\mathrm{O(I)}^{2-}$ is the energetically most favorable state.

We can therefore conclude that {\color{black} for $n$-type doping conditions} in both {\GO} and {\AO}, C$_\mathrm{O}$ acts as a compensating acceptor .
The high formation energy indicates it will not be a concern in {\GO}, but it could more easily incorporate in {\ALGO} alloys.  However, as evident from Fig.~\ref{fig:co}, this sensitively depends on chemical potentials, with cation-rich (oxygen-poor) conditions posing more of a problem.  We will return to this issue in Sec.~\ref{co-h}.

\text

\subsection{Hydrogen}\label{hresult}

\begin{figure}
\includegraphics[width=0.52\textwidth]{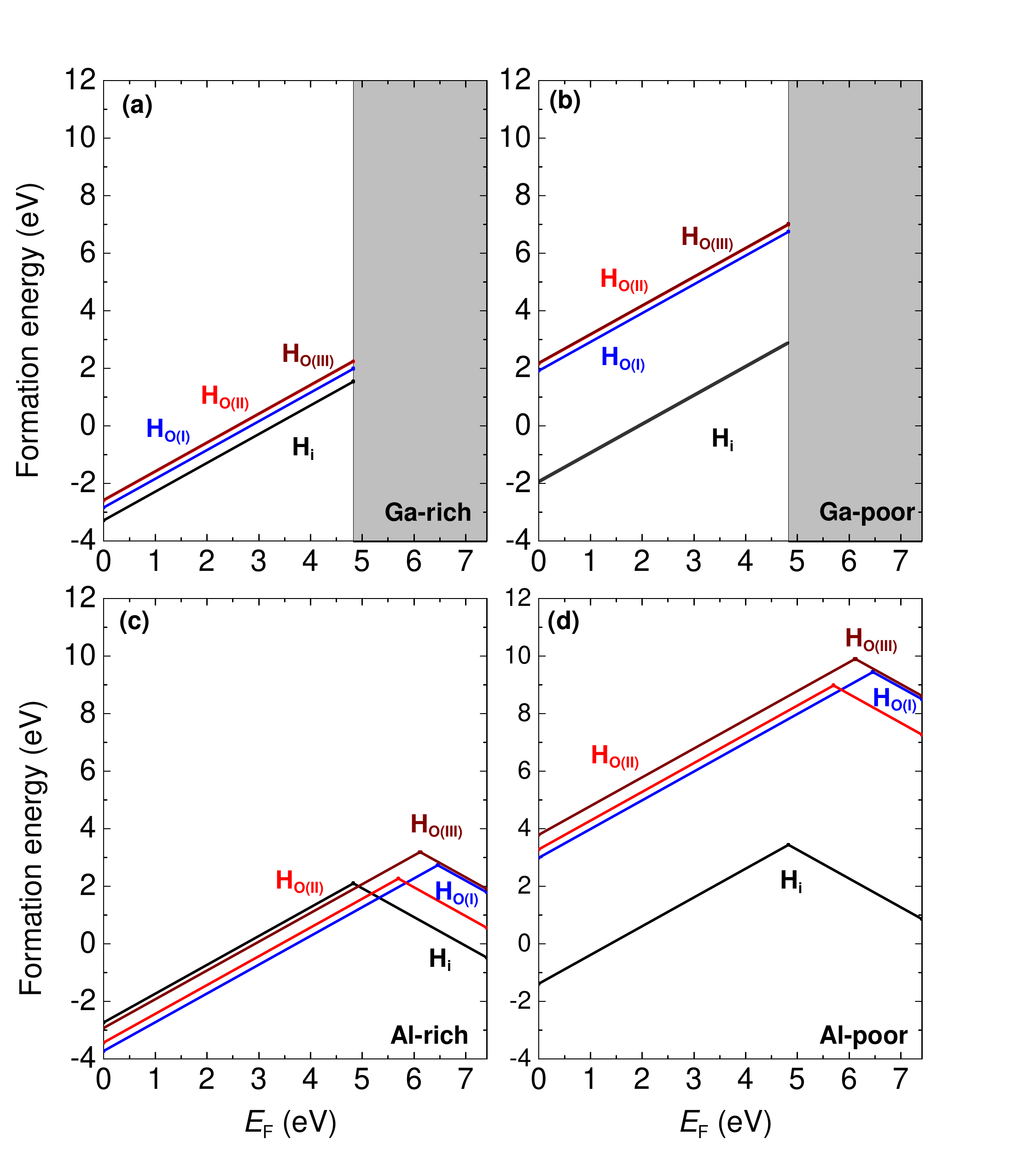}
\caption{\label{fig:H}
Formation energy versus Fermi level for substitutional and interstitial hydrogen in (a)-(b) {\BGO} and (c)-(d) {\TAO}. (a) and (c) are for cation-rich, and (b)-(d) for cation-poor conditions.
The grey area indicates the conduction band of {\BGO}.}
\end{figure}

Hydrogen is another impurity that is commonly unintentionally incorporated, particularly in MOCVD-grown samples.
We investigate H substituting on the O site (H$_\text{O(I)}$, H$_\text{O(II)}$, and H$_\text{O(III)}$) and H on the interstitial site (H$_i$) in {\TAO};
their formation energies are shown in Figs.~\ref{fig:H}(c,d). For comparison, we also show our calculated formation energies in {\BGO} [see Figs.~\ref{fig:H}(a,b)], which compare well with previous reports~\cite{varley2010oxygen}.
In {\BGO}, all these configurations were found to be shallow donors~\cite{varley2010oxygen}, but in {\TAO} they are stabilized in the negative charge state and thus act as acceptors when the Fermi level is high in the gap.

\begin{figure*}
\includegraphics[width=0.95\textwidth]{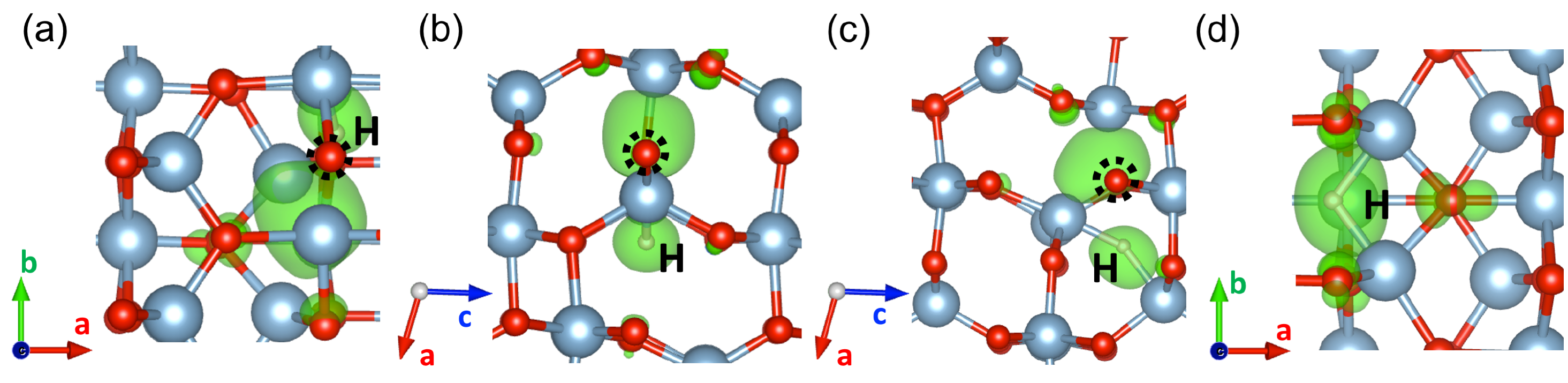}
\caption{\label{fig:h_local}\text
Local structure of (a) H$^-_\text{O(I)}$ (b) H$^-_\text{O(II)}$, (c) H$^-_\text{O(III)}$ and (d) H$^-_\text{i}$ in {\TAO}. Light blue, red and gray spheres denote Al, O and H atoms. The position of the O atom that was removed in H$^-_\text{O}$ was labeled using a black dashed circle. Isosurfaces of the charge density (0.01 $e$/\AA$^3$) of the localized electrons are shown in green.}
\end{figure*}

\subsubsection{Substitutional hydrogen, $\mathrm{H}_\mathrm{O}$}

In {\AO}, substitutional H$^+_\text{O}$ favors incorporation on the O(I) site.
Incorporation of a proton on the O(II) [O(III)] site is 0.29 [0.80 eV] higher in energy (see Fig.~\ref{fig:H}).
The same site preference is also observed in the ordered AlGaO$_3$ alloy: total energies of H$^+_\text{O(II)}$ and H$^+_\text{O(III)}$ are 0.08 eV and 0.33 eV higher than that of H$^+_\text{O(I)}$, respectively. In {\GO}, H$^+_\text{O(I)}$ is also most favorable, and H$^+_\text{O(II)}$ and H$^+_\text{O(III)}$ are slightly higher in energy \cite{varley2010oxygen}.

Under Al-rich conditions, H$_\text{O}$ has a low formation energy ($<$2 eV) in {\AO}, meaning it is easy to incorporate.
H$_\text{O(I)}$ and H$_\text{O(II)}$ behave as negative-$U$ centers with $U$=$-$0.10 eV and $-$0.16 eV, and a ($+/-$) charge-state transition level at 1.05 eV and 1.71 eV below the CBM (see Table~\ref{tab_sum} and Fig.~\ref{fig:H}). H$_\text{O(III)}$ is \textit{almost} a negative-$U$ center, with $U$=+0.03, and
($+/0$) and ($0/-$) charge-state transition levels at 6.11 eV and 6.13 eV, indicating a tiny (0.02 eV) range of stability of the neutral charge state.
The formation of negative-$U$ centers is due to the pronounced energy lowering in their negative charge state, induced by major local structural distortions.
The local structures are illustrated in Figs.~\ref{fig:h_local}(a--c);
two electrons are localized around H and nearby Al sites.
In the case of H$^-_\text{O(I)}$ [Fig.~\ref{fig:h_local}(a)], the H atom is still relatively close to the O(I) site, but the configurations of H$^-_\text{O(II)}$ [Fig.~\ref{fig:h_local}(b)] and H$^-_\text{O(III)}$ [Fig.~\ref{fig:h_local}(c)] are more like complexes of an oxygen vacancy plus a nearby interstitial H.

H$_\text{O(I)}$, H$_\text{O(II)}$, H$_\text{O(III)}$ will act as acceptors {\color{black} in $n$-type} {\ALGO} alloys due to the emergence of the ($+/-$) transition level in the band gap starting as a critical Al composition $x^\text{onset}$.
Interpolating ($+/-$) charge-state transition levels between {\GO} and the ordered AlGaO$_3$ alloy, the onset Al concentrations for H$_\text{O(I)}$, H$_\text{O(II)}$, H$_\text{O(III)}$ acting as compensating centers are 37\%, 13\% and 41\%, respectively (see Fig.~\ref{fig:band} and Table~\ref{tab_sum}).

\subsubsection{Interstitial hydrogen, $\mathrm{H}_i$}
\label{ssec:Hi}

Turning now to interstitial H in {\AO}, we find that the lowest-energy configuration in the neutral and positive charge state has H bonded to the three-fold coordinated O(I).
In the negative charge state, H$_i$ prefers to sit near two Al(I) atoms; the local geometry is illustrated in Fig.~\ref{fig:h_local}(d).
The local configurations of the $-$, $0$, and $+$ state of H$_i$ are similar to those in {\GO}~\cite{varley2010oxygen}.
In the neutral charge state, the Kohn-Sham state of H$_i$ lies about $\sim$ 3 eV below the CBM.
In the negative charge state, the Kohn-Sham state is about 1.3 eV above the VBM, as a result of the formation of two H-Al bonds.

As seen in Fig.~\ref{fig:H} and Table~\ref{tab_sum}, the ($+/-$) level of H$_i$ in {\AO} lies at 4.83 eV above the VBM.
H$_i$ is a negative-$U$ center, with a large magnitude of $U$=$-$1.99 eV.
Knowledge about the position of the ($+/-$) levels in {\BGO} and {\TAO} allows us to estimate the band lineup between these materials, according to Ref.~\onlinecite{van2003universal}; this leads to an alignment where the VBM of {\BGO} lies 0.01 eV below the VBM of {\TAO}.
We can compare this result with values obtained by alignment with the vacuum level derived from surface calculations; depending on the surface orientation, this procedure produces values ranging from +0.40 to $-$0.08 eV~\cite{mu2020orientation}.
The overall agreement with the hydrogen alignment is gratifying.

Figure~\ref{fig:H} shows that the formation energies of H$_i$ in {\AO} are quite low, suggesting that H$_i$ will readily incorporate, and in $n$-type material it will act as a compensating acceptor.
Following our interpolation procedure based on the case of {\GO} and ordered AlGaO$_3$ alloy, the ($+/-$) transition level of H$_i$ will already appear in the band gap of {\ALGO} alloys at an Al concentration of 1\% (see Fig.~\ref{fig:band}).

\subsubsection{Migration of interstitial hydrogen}

\begin{figure}
\includegraphics[width=0.48\textwidth]{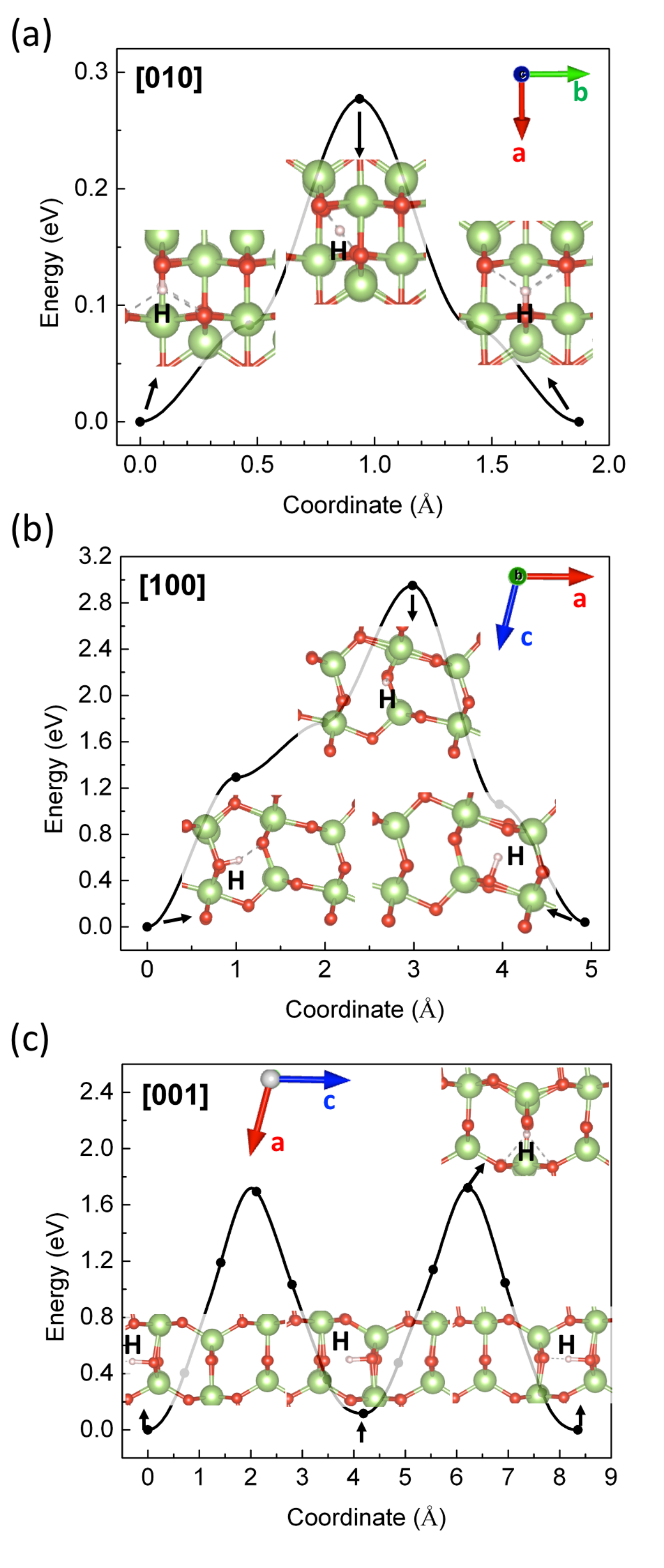}
\caption{\label{fig:h_p}  Potential energy for H$^+_i$ migration along the (a) [010], (b) [100], (c) [100] directions in {\BGO}.
Various configurations along the migration path are shown as insets.  Green and red spheres denote Ga and O atoms, respectively. Note the very different scales for the potential energy surfaces between (a) and (b) or (c).
}
\end{figure}

Since hydrogen is expected to be quite mobile, in order to assess whether hydrogen can play a role in compensation it is important to also assess migration barriers.
We investigated the migration barrier ($E_\text{b}$) of H$_i$ using the PBE functional and a one-shot HSE method, as described in Sec.~\ref{calc_detail}. The results are listed in Table~\ref{tab:eb}.  The annealing temperatures at which H becomes mobile for $+$ and $-$ charge states and for various crystallographic directions, calculated using Eq.~(\ref{anneal}), are also included in Table~\ref{tab:eb}.
As expected, the migration of H$_i$ in the monoclinic structure is anisotropic.
In {\GO}, we find the migration barrier of H$^+_i$ along the [010] direction to be 0.28 eV, in good agreement with the 0.34 eV value reported in Ref.~\onlinecite{varley2010oxygen}.  The migration path is illustrated in Fig.~\ref{fig:h_p}(a).
Migration along other directions has higher barriers.
Along [100] [Fig.~\ref{fig:h_p}(b)] the barrier is as high as 2.95 eV, and along [001] Fig.~\ref{fig:h_p}(c), the barrier is 1.73 eV.
The annealing temperature for H$^+_i$ migration along [010], [100] and [001] directions are
101, 1062, and 623 K, respectively.

\begin{table}
\caption{Calculated migration barriers ($E_\text{b}$, eV) of H$^+$ and H$^-$ in {\GO} and {\AO} along three crystallographic directions. The calculated annealing temperature ($T_\text{anneal}$, K) is listed, assuming a hopping rate of 1 $s^{-1}$.
}
\begin{ruledtabular}
\begin{tabular}{c|ccc|ccc}
            & \multicolumn{3}{c|}{H$^+$} & \multicolumn{3}{c}{H$^-$} \\
  \hline
{\GO} & [010] & [100] & 	[001]	& [010]	& [100]	& [001] \\
\hline
$E_\text{b}$ 	& 0.28 & 	2.95 &  1.73 & 	2.49	& 1.83 & 	1.57 \\
$T_\text{anneal}$	& 101 & 	1062 & 	623 & 	896 & 	659 & 	565  \\
\hline
{\AO} & \multicolumn{3}{c|}{ }& \multicolumn{3}{c}{ } \\
\hline
$E_\text{b}$ &	0.23 &	2.29	&	1.78 &	1.33	& 2.67 &	1.83 \\
$T_\text{anneal}$ &	83&	824	&641 	& 479 &	961	& 659 \\
\end{tabular}
\end{ruledtabular}
\label{tab:eb}
\end{table}

Our calculated results for migration paths and barriers for H$^+_i$ in {\AO} are quite similar to the results in {\GO}; migration is also strongly anisotropic (see Table~\ref{tab:eb}).
The low migration barrier suggests that H$^+_i$ can move along the [010] direction (which is the most commonly used growth orientation) even at temperatures well below room temperature.

For H$^-_i$, the migration barriers are generally higher.
Particularly striking is that migration in {\GO} now has the highest barrier along [010] (see Table~\ref{tab:eb}).
We attribute this to the fact that in the barrier geometry for H$^-_i$, the distance between H and the nearest Ga atoms is quite large (2.1 {\AA}).  Since it is Coulombic attraction between H$^-_i$ and the cations that stabilizes the structure, the lack of strong bonding at the saddle point raises the energy.
In contrast, in {\AO} the corresponding distance is only 1.8 {\AA}, providing for a stronger interaction with the Al atoms and lowering the barrier height.
While the migration barriers for H$^-_i$ are higher than for H$^+_i$, the annealing temperatures in Table~\ref{tab:eb} indicate that H$^-_i$ may be mobile at temperatures at which growth or processing of {\GO} and {\ALGO} alloys is typically performed.

Figuring out precise migration barriers for H$_i$ in {\ALGO} alloys may not be as simple as interpolating the values in {\GO} and {\AO}.  The barriers to be overcome will be determined by the local environment.  If, at a particular Al composition, a ``percolation path'' exists that allows hydrogen to follow favorable atomic arrangements, the lower of the two migration barriers will likely apply.  In the absence of such a path, the higher of the two barriers will need to be overcome.

We note that the mobility of H$^-_i$ will be reduced by binding to positively charged impurities, in particular donor dopants.  Formation of a Si--H complex will increase the stability of H$^-_i$ in the lattice and add to the overall activation energy for hydrogen motion.
Therefore interstitial H$^-_i$, in addition to substitutional H$^-_\text{O}$, may play a role in compensation of shallow donors in {\ALGO}.

\subsection{Complexes}\label{comp}

Since hydrogen can easily incorporate and has relatively low migration barriers, we need to assess the possibility of complex formation between hydrogen and other impurities, either with the intentionally incorporated Si donor, or with unintentional impurities such as C.

\subsubsection{Si$_\mathrm{cation}$--H complexes} \label{si-h}

\begin{figure}
\includegraphics[width=0.52\textwidth]{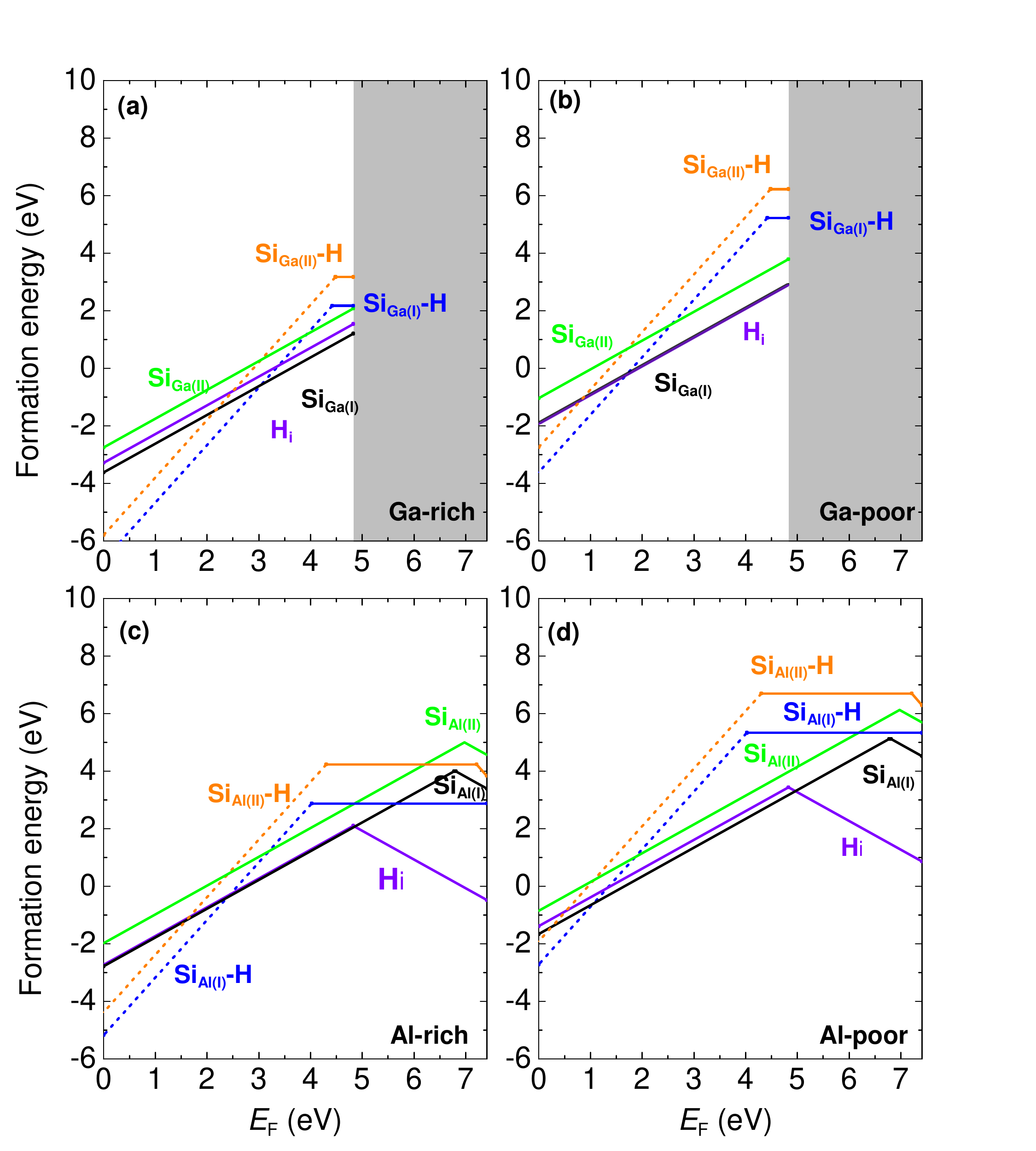}
\caption{\label{fig:si-h}
Formation energy versus Fermi level for isolated Si impurities, H interstitials, and Si--H defect complexes in (a)-(b) {\BGO} and (c)-(d) {\TAO}. (a) and (c) are for cation-rich, and (b)-(d) for cation-poor conditions.
Dashed lines denote thermodynamic instability of the defect complex, as explained in the text.
The grey area indicates the conduction band of {\BGO}.
}
\end{figure}

We first discuss complexes with Si on the cation site;
as discussed in Sec.~\ref{siresult}, we found that Si$_\mathrm{cation}$ is an effective donor in {\ALGO} alloys up to 70\% Al incorporation for Si$_\mathrm{(I)}$ and 81\% Al incorporation for Si$_\mathrm{(II)}$.
The formation energy of Si--H complexes in {\GO} is shown in Fig.~\ref{fig:si-h}(a) for Ga-rich and Fig.~\ref{fig:si-h}(b) for Ga-poor conditions.
We find that similar to the Si$_\mathrm{Ga}$ impurity, the Si$_\mathrm{Ga}$--H complex also has lower energy when Si occupies the tetrahedral cation (I) site.

The formation energy of Si--H complexes under Ga-poor conditions is simply a rigid shift to higher energies of the formation energy under Ga-rich conditions [Figs.~\ref{fig:si-h}(a,b)]; our discussion about the properties of complexes applies to both.
We find that when the Fermi level is high in the gap, Si$_\mathrm{Ga(I)}$--H complexes are stable in the neutral charge state, indicating that the Si donor has been passivated.
We also find that Si$_\mathrm{Ga(I)}$--H has a (2$+/0$) charge-state transition level at 0.41 eV below the CBM.
The formation of a Si--H complex might seem unexpected, given that the constituents, Si$_\mathrm{Ga}$ and H$_i$, both prefer to occur in the positive charge state throughout the band gap of {\GO}, as evident from Fig.~\ref{fig:si-h}(a).  Our calculated formation energy of point defect Si$_\mathrm{Ga}$ and H$_i$ in {\GO} agree well with the previous study~\cite{varley2010oxygen}.
However, as seen in Table~\ref{tab_sum}, the ($+/-$) level of H$_i$ in {\GO} lies at 4.84 eV above the VBM, i.e., just 0.01 eV above the CBM.  It is therefore to be expected that for Fermi levels near the CBM, H$_i^-$ could be stable enough to lead to formation of a complex with  Si$_\mathrm{Ga}^+$, particularly when Coulomb attraction is taken into account.

This is confirmed by evaluation of the binding energy of the complex, defined as
\begin{equation}\label{eq:bind}
\begin{split}
E_\mathrm{bind}[(\mathrm{Si}_\mathrm{Ga(I)}-\mathrm{H})^0] = E^f(\mathrm{Si}_\mathrm{Ga(I)}^+) +E^f(\mathrm{H}_i^-) \\
- E^f[(\mathrm{Si}_\mathrm{Ga(I)}-\mathrm{H})^{0}]
\end {split}
\end{equation}
where a positive value of the binding energy signifies a stable, bound complex.
This yields a binding energy of 0.59 eV.
This value (which is smaller than the formation energy of each of the constituents) indicates the complex will not be stable during growth at high temperatures, since entropy favors the isolated entities (see Ref.~\onlinecite{jap04}).  In addition, formation after growth is not very likely since H$_i$ will predominantly occur in the positive charge state, which is repelled by Si$_\mathrm{Ga}^+$.

One may then wonder about the stability of (Si$_\mathrm{Ga(I)}$--H)$^{2+}$.
Indeed, a binding-energy calculation in this case yields a negative value for E$_\mathrm{bind}$ ($-$0.24 eV), i.e., the sum of the formation energy of the constituents, Si$_\mathrm{Ga(I)}^+$ and H$_i^+$, is smaller than the formation energy of the (Si$_\mathrm{Ga(I)}$--H)$^{2+}$ complex.
We show the 2+ charge state of this complex with dashed lines in Fig.~\ref{fig:si-h}, to indicate that this is a locally stable configuration that is, however, thermodynamically unstable.

\begin{figure*}
\includegraphics[width=0.98\textwidth]{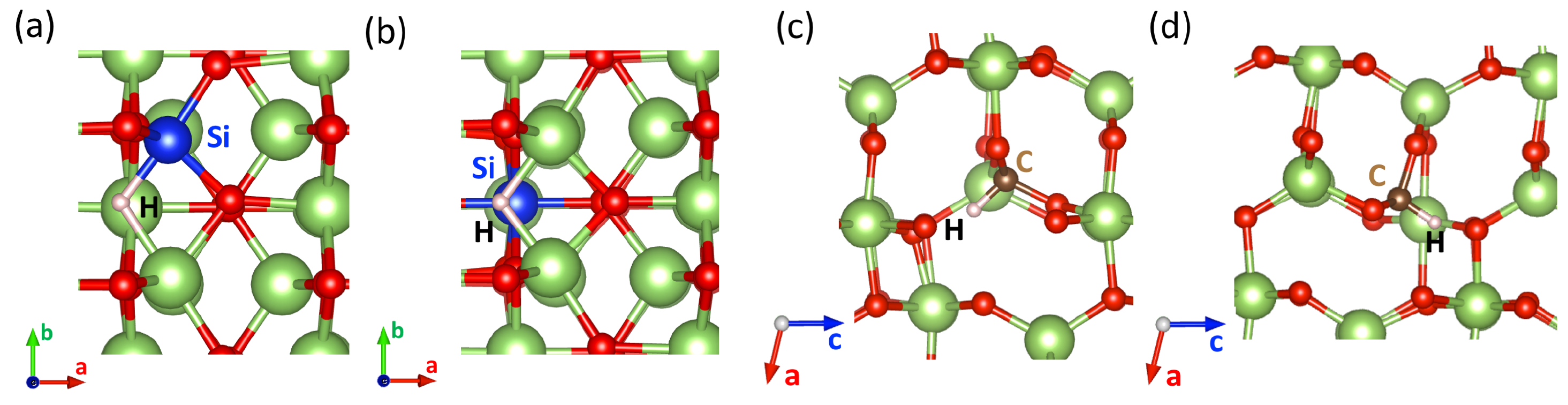}
\caption{\label{fig:complex_geom}\text
Local structure of (a) (Si$_\mathrm{Ga(I)}$--H)$^0$ (b) (Si$_\mathrm{Ga(II)}$--H)$^0$, (c) (C$_\mathrm{Ga(I)}$--H)$^0$ and (d) (C$_\mathrm{Ga(II)}$--H)$^0$ in {\GO}. Green, red, gray, blue and brown spheres denote Al, O, H, Si and C atoms, respectively.}
\end{figure*}

The notion that (Si$_\mathrm{Ga(I)}$--H)$^0$ results from bonding between Si$_\mathrm{Ga(I)}^+$ and H$_i^-$ is consistent with the local geometry of the complex:
we found in Sec.~\ref{ssec:Hi} that H$_i^{-}$ prefers to bind to two cation atoms on tetrahedral sites, and the configuration of (Si$_\mathrm{Ga(I)}$--H)$^0$ is such that the hydrogen atom is located close to a Ga(I) and to the Si(I) atom as seen from Fig.~\ref{fig:complex_geom}(a). This is very similar to the geometry shown in Fig. ~\ref{fig:h_local}(d). For (Si$_\mathrm{Ga(I)}$--H)$^0$, the bond length of Si(I)-H and Ga(I)-H are 1.53 \AA\ and 1.77 \AA, respectively.

If the (Si$_\mathrm{Ga(I)}$--H)$^{0}$ complex does form during cooldown, one may wonder about its stability at room temperature.
The activation energy for dissociation can be estimated by adding the migration barrier for the mobile hydrogen species to the binding energy of the complex.
All of the H$_i^-$ migration barriers are large enough (Table~\ref{tab:eb}) to make this activation energy sufficiently high and keep the complex stable at room temperature.  Using the lowest migration barrier for H$_i^-$ in {\GO} in Table~\ref{tab:eb}, we would estimate an activation energy of 0.59+1.57=2.16 eV, which would correspond to an annealing temperature of 778 K.
However, since the (2+/0) transition level is so close to the CBM, an alternative dissociation mechanism could occur in which electron excitation to the CBM converts the complex to a 2+ charge state, after which the dissociation will proceed more easily both because of the lower migration barrier of H$_i^+$ (Table~\ref{tab:eb}) and because of the Coulomb repulsion between H$_i^+$ and Si$_\mathrm{Ga}^+$.

Venzie {\it et al.}~\cite{venzie2021oh} recently reported that exposure of Si-doped {\GO} to a hydrogen plasma led to passivation of the Si donor, as evidenced by a reduction in conductivity and an increase in mobility.  The mobility increase indicates that the concentration of ionized-impurity centers is reduced, consistent with the formation of a complex (as opposed to mere compensation, where the compensating centers are spatially separated from the ionized donors).
Venzie \textit{et al.}~\cite{venzie2021oh} also proposed a geometry for the complex, which suggests that H binds to an O(I) neighbor of the Si donor; our calculations show that this configuration is 3.23 eV higher in energy than the stable geometry where H is bonded to two cations [see Fig.~\ref{fig:complex_geom}(a)].
The configuration with cation-H bonds leads to vibrational frequencies significantly lower than the 3477.6 cm$^{-1}$ attributed to the Si-H complexes\cite{venzie2021oh}, suggesting that these observed modes may be due to other types of complexes that form upon hydrogenation.

The formation energy of the Si$_\mathrm{Ga(II)}$--H complex is higher, but otherwise its characteristics are very similar to those of the Si$_\mathrm{Ga(I)}$--H complex. The (2$+/0$) level of Si$_\mathrm{Ga(II)}$--H in {\GO} occurs at 0.35 eV below the CBM (compared to 0.41 eV for Si$_\mathrm{Ga(I)}$--H),
and the binding energy is 0.47 eV.  The local geometry of (Si$_\mathrm{Ga(II)}$--H)$^0$ is similar to that of (Si$_\mathrm{Ga(I)}$--H)$^0$, with a hydrogen binding to two nearby tetrahedral cation sites [see Fig.~\ref{fig:complex_geom}(b)]. The H-Ga bond lengths are both 1.67 \AA. This reinforces the argument that a (Si$_\mathrm{Ga}$--H)$^0$ complex in {\GO} is related to the H$_i^-$ point defect.

Similar complexes form in {\AO}, as seen in Fig.~\ref{fig:si-h}(c).
The (2$+/0$) level of the (Si$_\mathrm{Al(I)}$--H) complex occurs at $E_\mathrm{F}$ = 4.02 eV (3.39 eV below the CBM), and the
binding energy of the complex is E$_\mathrm{bind}$[(Si$_\mathrm{Al(I)}$-H)$^0$]=1.27 eV.
For Si$_\mathrm{Al(II)}$--H, we find the (2$+/0$) transition level at 4.31 eV above the VBM (3.10 eV below the CBM), with a
binding energy E$_\mathrm{bind}$[(Si$_\mathrm{Al(I)}$--H)$^0$]=0.71 eV.

Based on our calculated numbers, we can assess whether complex formation with hydrogen has an impact on conductivity in Si-doped {\ALGO} alloys.
We have found the neutral complex to be stable in both {\GO} and {\AO} for Fermi levels close to the CBM, and hence we expect the complex to also be stable in the alloys, with a binding energy that increases with increasing Al content.
Indeed, the experiments of Venzie {\it et al.}~\cite{venzie2021oh} indicated that intentional hydrogenation can produce the complex in {\GO}.
However, as discussed above, the relatively modest binding energy indicates the complex will not be formed at the growth temperature.
Complex formation would therefore need to occur during cool down.  Since this requires bringing H$_i$ close to a Si impurity in the positive charge state, the increasing stability of H$_i^-$ as the Al concentration increases makes complex formation more likely.
If complex formation would indeed occur, annealing the sample would serve to dissociate the complex and activate the Si donor, as is well known for hydrogen-related complexes in other semiconductors~\cite{AR2006}.

\subsubsection{Si$_\mathrm{O}$ and Si$_\mathrm{O}$--H complexes} \label{sio-h}

For completeness, we also investigated Si$_\mathrm{O}$ and Si$_\mathrm{O}$--H complexes in both {\GO} and {\AO}.
Formation energies are shown in the Supplemental Material (SM) \cite{SM}.
Si$_\mathrm{O}$ itself is stable in positive charge states when the Fermi level is low, but for Fermi levels close to the CBM
Si$_\mathrm{O(I)}$ and Si$_\mathrm{O(III)}$ behave as compensating acceptors, while Si$_\mathrm{O(II)}$ is neutral.
However, the formation energies are so large in $n$-type material (even under the most favorable O-poor conditions) that they are
unlikely to form or play a role in carrier compensation.
Si$_\mathrm{O}$--H complexes also act as acceptors when $E_\text{F}$ is high in the gap, but again their formation energies are so large that they are
unlikely to form.

\subsubsection{C$_\mathrm{cation}$--H complexes} \label{c-h}

\begin{figure}
\includegraphics[width=0.52\textwidth]{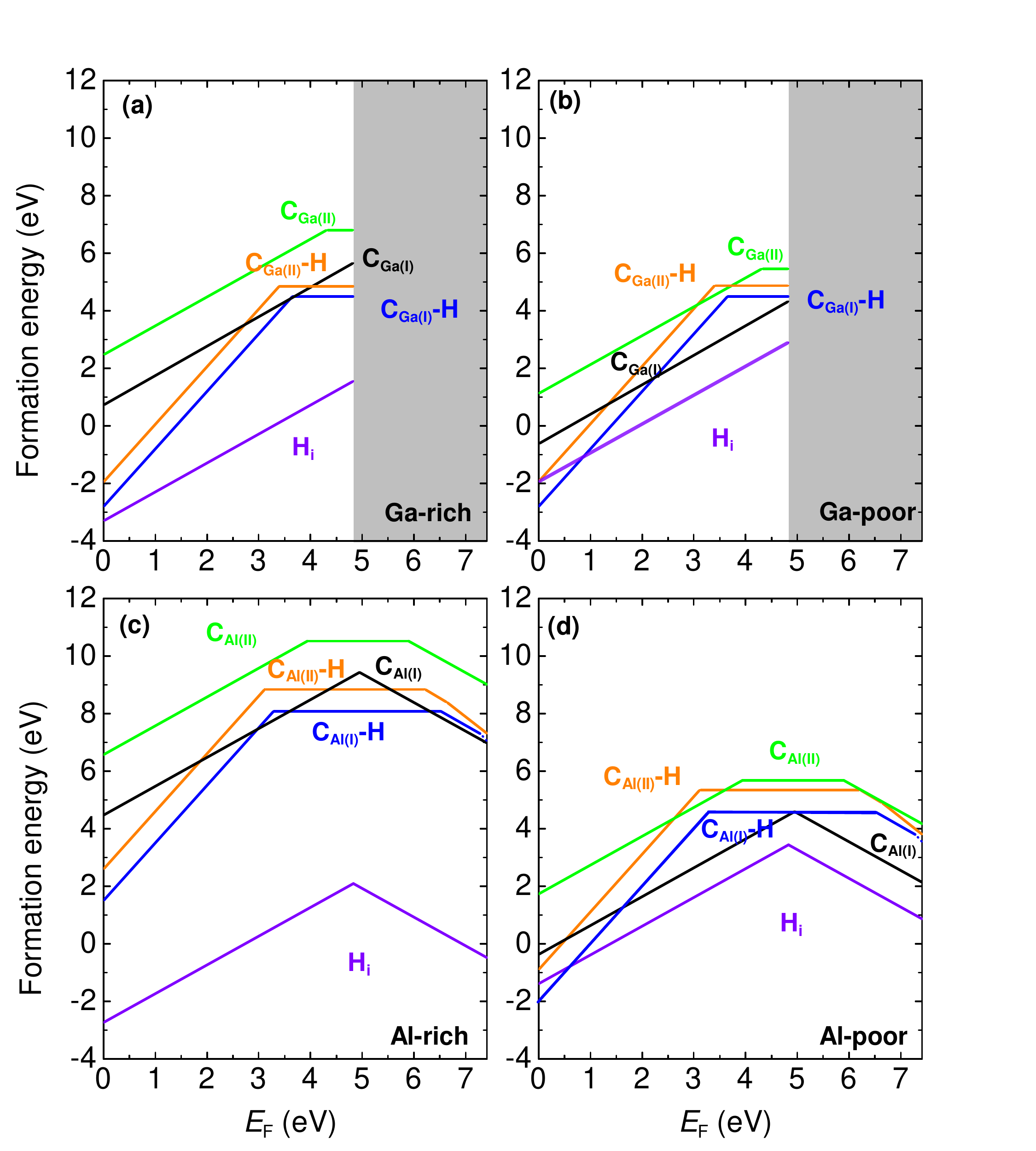}
\caption{\label{fig:c-h}
Formation energy versus Fermi level for isolated C$_\mathrm{cation}$ impurities, H interstitials, and C$_\mathrm{cation}$--H defect complexes in (a)-(b) {\BGO} and (c)-(d) {\TAO}. (a) and (c) are for cation-rich, and (b)-(d) for cation-poor conditions.
The grey area indicates the conduction band of {\BGO}.
}
\end{figure}

Figures \ref{fig:c-h}(a,b) show the formation energies of C--H complexes in {\GO}.
Similar to the Si$_\mathrm{Ga(I)}$-H in {\GO}, a (2+/0) charge-state transition level of C$_\mathrm{Ga(I)}$-H occurs in the band gap, at 1.18 eV below the CBM, despite the fact that the individual impurities C$_\mathrm{Ga(I)}$ and H$_i$ are single donors and occur in the positive charge state throughout the band gap.
Following a similar definition as in Eq.~(\ref{eq:bind}), we can define a binding energy relative to C$_\mathrm{Ga(I)}^+$ and H$_i^-$, finding E$_\mathrm{bind}$[(C$_\mathrm{Ga(I)}$--H)$^0$]=2.62 eV.
This much greater value of the binding energy and lower position of the (2+/0) transition level for the C$_\mathrm{Ga(I)}$--H complex as compared to the Si$_\mathrm{Ga(I)}$--H complex indicates that a different type of bonding occurs.
As pointed out in Sec.~\ref{si-h}, (Si$_\mathrm{Ga(I)}$--H)$^0$ can be viewed as a combination of Si$^+$ and H$^-$, with the H atom bonding to two tetrahedral cation sites [Fig.~\ref{fig:complex_geom}(a)].
In (C$_\mathrm{Ga(I)}$--H)$^0$, on the other hand, the H atom binds to C within the tetrahedral oxygen cage, forming a strong C-H bond with a bond length of 1.06 {\AA} [see Fig.~\ref{fig:complex_geom}(c)].
The strength of this bond, which is related to the small atomic size of C, makes the C-H combination behave as a unit, similar to a nitrogen atom, and lowers the overall formation energy.
For the C$_\mathrm{Ga(I)}$--H complex, the 2+ charge state is actually thermodynamically stable, i.e., it is 0.24 eV lower in energy than the sum of the formation energies of C$^+_\mathrm{Ga(I)}$ and H$_i^+$.
In the donor configuration of the C$_\mathrm{Ga(I)}$--H complex, H is bonded to a neighboring oxygen atom within the tetrahedral oxygen cage instead of to C.
This is also true for other donor (i.e., positively charged) configurations of the C$_\mathrm{cation}$--H complex in both {\GO} and {\AO}.

Similar behavior is observed for C$_\mathrm{Ga(II)}$--H: a (2+/0) transition level occurs at 2.25 eV below the CBM, and the binding energy of the neutral complex is 4.00 eV.
In this case,  the 2+ charge state is also thermodynamically stable, i.e., it is 1.13 eV lower in energy than the sum of the formation energies of C$^+_\mathrm{Ga(II)}$ and H$_i^+$ [see Figs.~\ref{fig:c-h}(a,b)].
Again, a C-H bond with a bond length of 1.06 \AA\ forms within the octahedral oxygen cage for (C$_\mathrm{Ga(II)}$--H)$^0$ [see Fig.~\ref{fig:complex_geom}(d)]. This C--H bond formation thus indicates a unique structural feature of the C--H complex regardless of which cation site C substitutes on.

Figure~\ref{fig:c-h}(a) shows that C$_\mathrm{Ga}$--H complexes are lower in energy than isolated C$_\mathrm{Ga}$ impurities under $n$-type doping conditions,
and this is true over most of the range of Ga chemical potentials.
In addition, because of their high binding energy, such complexes may well be present during growth.
Measurements of carbon concentrations in MOCVD-grown samples, in which both carbon and hydrogen are likely to be unintentionally incorporated, may thus reflect the presence of such complexes rather than isolated C$_\mathrm{Ga}$ donors.
Since the complexes are neutral, they would not impact the conductivity of the sample.
This implicit passivation of carbon impurities, which would otherwise act as shallow donors,
may well explain why MOCVD-grown {\GO} samples can exhibit free-carrier concentrations as low as 10$^{14}$ cm$^{-3}$~\cite{alema2020}.

We now turn to  C$_\mathrm{Al}$--H complexes in {\AO}.
Figures~\ref{fig:c-h}(c,d) show the formation energy of the complexes in {\AO} in Al-rich conditions.
The type of bonding described for {\GO} is also present in {\AO}, but now we see that for $E_\text{F}$ high in the gap also a negative charge state can be stabilized, which indicates that the complex could potentially act as an acceptor and cause compensation.
However, we note that the (0/$-$) transition level of the complex occurs at a higher position in the band gap than the (+/$-$) transition of the isolated C$_\mathrm{Al}$.
The ($0/-$) level of C$_\mathrm{Al(I)}$--H occurs at 0.90 eV below the CBM, and
a ($-/2-$) level is also present, at 0.13 eV below the CBM.
Similarly, ($0/-$) and ($-/2-$) levels of C$_\mathrm{Al(II)}$--H are also observed close to the CBM, at 1.20 and 0.33 eV below the CBM.
We note that the formation energy for C$_\mathrm{Al(II)}$--H in monoclinic {\AO} is similar to that in \textit{corundum} {\AO}, except that the 1$-$ charge state of C$_\mathrm{Al(II)}$--H is absent in corundum~\cite{choi2014hydrogen}.  The C$_\mathrm{Al(II)}$--H bonding in monoclinic {\AO} is also very similar to that in the corundum phase~\cite{choi2014hydrogen}, implying that the C-H bonding within an octahedral O cage [as in Fig.~\ref{fig:complex_geom}(d)] is a host-independent structural feature.

We verified that the C$_\mathrm{Al}$--H complexes are thermodynamically stable (i.e., have positive binding energies) in both the neutral and negative charge states. The only exception is (C$_\mathrm{Al(I)}$--H)$^{2-}$, which has a negative binding energy.  Formation of this complex is highly unlikely, anyway, since it would require the Fermi level to be within 0.13 eV of the CBM.

The stability of the C$_\mathrm{cation}$--H complexes in the neutral and negative charge states in both {\GO} and {\AO} implies they will also be stable in {\ALGO} alloys.
The question is whether the behavior as a compensating acceptor that we find in {\AO} can occur in {\ALGO} alloys.
To assess this, we calculated the position of the (0/$-$) transition level in {\GO} and combined this information with the results for the (0/$-$) level in {\AO}.
This allows us to conclude that the C$_\mathrm{cation}$--H complexes can only occur in a negative charge state (i.e., act as acceptors) when the Al concentration exceeds 56\% for C$_\mathrm{(I)}$--H or 64\% for C$_\mathrm{(II)}$--H.

\subsubsection{C$_\mathrm{O}$--H complexes} \label{co-h}

\begin{figure}
\includegraphics[width=0.52\textwidth]{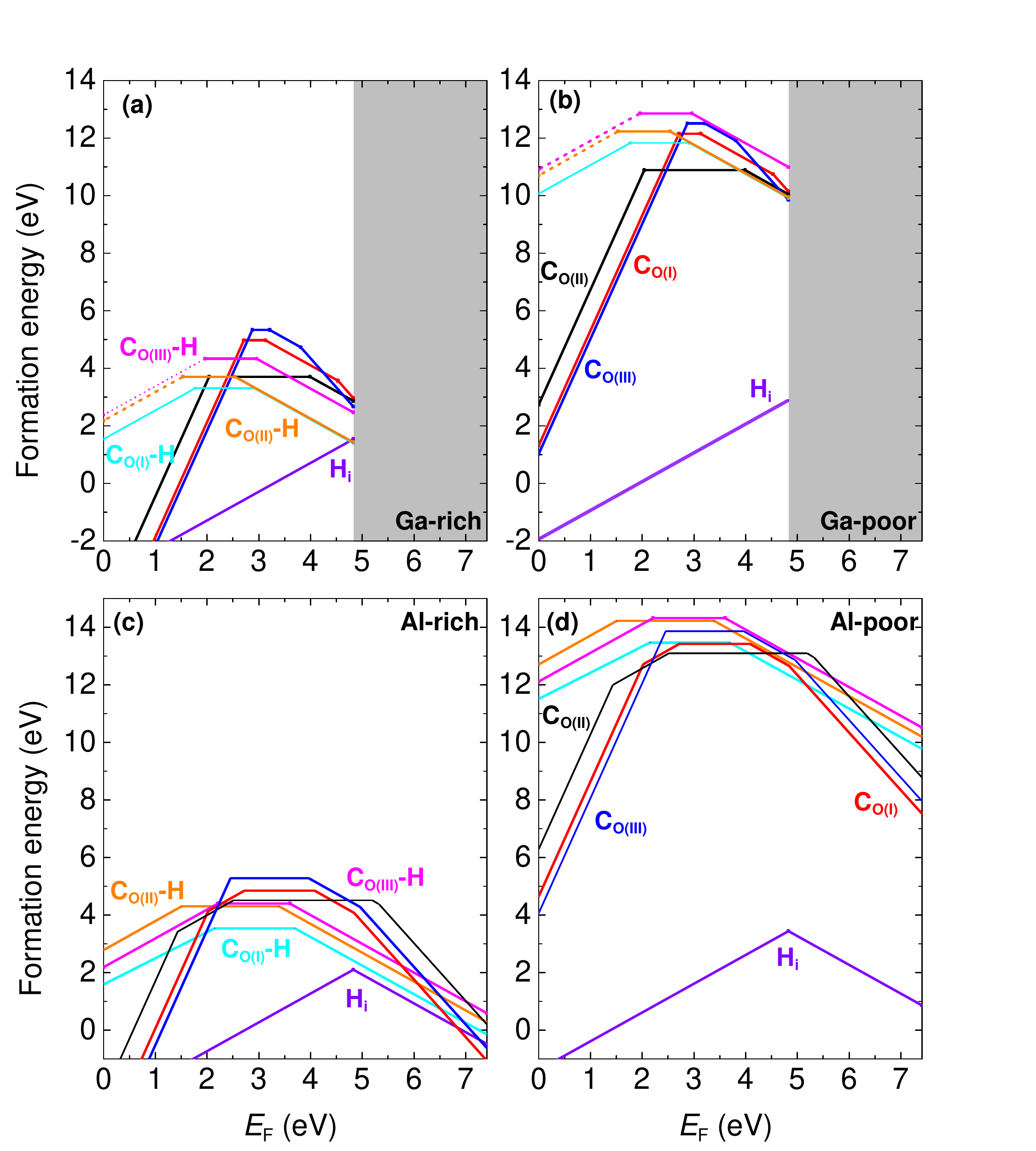}
\caption{\label{fig:co-h}
Formation energy versus Fermi level for C$_\mathrm{O}$ impurities, H interstitials, and C$_\mathrm{O}$--H defect complexes in (a)-(b) {\BGO} and (c)-(d) {\TAO}. (a) and (c) are for cation-rich, and (b)-(d) for cation-poor conditions.
Dashed lines denote thermodynamic instability of the defect complex, as explained in the text.
The grey area indicates the conduction band of {\BGO}.
}
\end{figure}

We discussed carbon on the oxygen site, C$_\mathrm{O}$, in Sec.~\ref{sec:co}; now we turn to its complexes with hydrogen.
As shown in {\color{black}Fig.~\ref{fig:co-h}(a)}, C$_\mathrm{O}$--H has very low formation energy in {\GO}, particularly when the Fermi level is high in the gap and under Ga-rich conditions.
Regardless of the O site [O(I), O(II), O(III)] on which carbon is substituting, ($+/0$) and ($0/-$) charge-state transition levels are present in the band gap.
We verified that higher negative charge states (such as 2$-$) are not stable for Fermi levels in the band gap.
In all these complexes, hydrogen is always bonded to the C atom, with a bond length between 1.07 \AA\ and 1.09 \AA.
Again, we note that the C-H unit behaves very similarly to a N impurity: C$_\mathrm{O}$-H and substitutional N$_\mathrm{O}$ have similar ($+/0$) and ($0/-$) transition levels in the gap, and both behave as acceptors in $n$-type material~\cite{peelaers2019deep}.

The behavior of C$_\mathrm{O}$--H in {\AO} is very similar: it also only exhibits ($+/0$) and ($0/-$) levels in the gap, and the complex has low formation energies under Al-rich conditions and when the Fermi level is high in the gap.
The binding energies of the (C$_\mathrm{O}$--H)$^{-}$ complexes are all large in both {\GO} and {\AO}: all binding energies exceed 3 eV, with the exception of the (C$_\mathrm{O(III)}$--H)$^{-}$ complex, where the binding energy is 1.76~eV.

We propose that the C$_\mathrm{O}$--H complexes may have a distinct impact on MOCVD-grown {\GO}.
Seryogin {\it et al.}~\cite{seryogin2020mocvd} grew {\GO} using a trimethylgallium (TMGa) precursor, which is known to potentially lead to increased C incorporation.
They found that films grown with a lower O$_2$/TMGa ratio were significantly more resistive and also contained significantly more carbon.
These results indicate that carbon is behaving as a compensating acceptor.
The authors of Ref.~\onlinecite{seryogin2020mocvd} pointed to a computational study~\cite{lany2018defect} that had found C$_\mathrm{Ga}$ to be a $DX$ center in {\GO}, with a ($+/-$) level 0.81 eV below the CBM.  That result is very different from ours, probably due to the particulars of the band-gap correction used in Ref.~\onlinecite{lany2018defect}.
In any case, if C$_\mathrm{Ga}$ is the culprit, moving from O-rich to O-poor conditions should reduce the C$_\mathrm{Ga}$ incorporation, which is not consistent with the observed increase in C concentration.
We think it is much more likely that C$_\mathrm{O}$ is involved, which indeed acts as a compensating acceptor with a concentration that should increase under O-poor conditions, as seen in Fig.~\ref{fig:co}.
However, that figure also shows that the formation energy of C$_\mathrm{O}$ in the negative charge state is still relatively high, even under the most favorable (extreme O-poor and C-rich) conditions.
{\color{black}Figure~\ref{fig:co-h} indicates} that this energy can be lowered by complexing with hydrogen, which should of course also be abundantly available during MOCVD growth.

The involvement of C$_\mathrm{O}$--H complexes is also consistent with the experiments of Alema {\it et al.}~\cite{alema2020h2o}, who found that adding H$_2$O vapor to the oxygen source decreased the carrier concentration, both in unintentionally doped (UID) and lightly Si-doped samples.
Hydrogen could either passivate donors or form compensating acceptors.
Donor passivation may occur by forming complexes with the Si donor (though in Sec.~\ref{si-h} we noted this was not very likely) or with unintentional C$_\mathrm{Ga}$ donors; for compensating acceptors, C$_\mathrm{O}$--H complexes are most likely, and indeed, Ref.~\onlinecite{alema2020h2o} reported an increase in compensating acceptors rather than a decrease in donor concentration upon adding H$_2$O.

\section{Conclusions} \label{conc}

We have reported a comprehensive investigation of silicon donors in {\ALGO} and the various ways in which they may be compensated.
For Si in {\TAO} we performed a detailed study of \textit{DX} center formation. Two stable \textit{DX} Si$^-_\text{Al(I)}$ configurations were identified, with the configuration involving bonding to two cations and no broken bonds being most energetically favorable.
Interpolating the ($+/-$) charge-state transition levels between {\BGO} and {\TAO}, we find that Si is an effective donor over a wide range of Al concentrations in {\ALGO} alloys, up to 70\% Al.

We also investigated the behavior of carbon and hydrogen impurities, which are commonly unintentionally present, particularly in MOCVD.
Even though these act as shallow donors in {\BGO}, they become deep centers in {\ALGO} alloys.
Based on interpolation of the ($+/-$) transition levels, we find that C$_\text{(I)}$ acts as a compensating acceptor already at 5\% Al, and H$_i$ already at 1\% Al.

Additionally, complex formation between hydrogen and either Si or C may occur.
We found that Si$_\mathrm{cation}$--H complexes have relatively low binding energy; if they form, they can probably quite easily be removed by annealing.
C$_\mathrm{Ga}$--H complexes are very stable and electrically neutral in $n$-type {\GO}, and may explain why unintentionally incorporated carbon does not affect the carrier concentration.\cite{alema2020}
C$_\mathrm{cation}$--H complexes do act as acceptors in {\ALGO} alloys, but only if the Al concentration exceeds 56\%.
C$_\mathrm{O}$--H, finally, was found to act as an acceptor over the entire range of alloy compositions, behaving very similarly to a N$_\mathrm{O}$ substitutional impurity
and explaining experimental observations of carbon-related compensation in Ga$_2$O$_3$ grown by MOCVD~\cite{seryogin2020mocvd,alema2020h2o}.

Our findings indicate that the presence of hydrogen or carbon may interfere with controlled low-level Si doping in {\ALGO} alloys, and our detailed results can be used to devise growth or processing conditions to avoid this.

\section*{Acknowledgments}
The authors acknowledge Hongping Zhao for fruitful discussions and preliminary experimental results.
S. M. is much indebted to Andrew J. E. Rowberg for stimulating discussions and valuable comments on the manuscript.
The work was supported by the GAME MURI of the Air Force Office of Scientific Research (FA9550-18-1-0479).
This work was partially performed under the auspices of the U.S. DOE by Lawrence Livermore National Laboratory (LLNL) under contract DE-AC52-07NA27344 and partially supported by LLNL Laboratory Directed Research and Development funding under project number 22-SI-003 and by the Critical Materials Institute, an Energy Innovation Hub funded by the U.S. DOE, Office of Energy Efficiency and Renewable Energy, Advanced Manufacturing Office.
Work at NRL was supported by the Office of Naval Research through the Naval Research Laboratory's Basic Research Program, and made use of DoD HPCMP resources.
Use was made of computational facilities purchased with funds from the National Science Foundation (NSF) (CNS-1725797) and administered by the Center for Scientific Computing (CSC). The CSC is supported by the California NanoSystems Institute and the Materials Research Science and Engineering Center (MRSEC; NSF DMR 1720256) at UC Santa Barbara.
This work also used the Extreme Science and Engineering Discovery Environment (XSEDE), which is supported by the National Science Foundation under Grant No. ACI-1548562, and
the Frontera resources at the Texas Advanced Computing Center (TACC) at The University of Texas at Austin (NSF OAC-1818253).


\begin{thebibliography}{70}%
\makeatletter
\providecommand \@ifxundefined [1]{%
 \@ifx{#1\undefined}
}%
\providecommand \@ifnum [1]{%
 \ifnum #1\expandafter \@firstoftwo
 \else \expandafter \@secondoftwo
 \fi
}%
\providecommand \@ifx [1]{%
 \ifx #1\expandafter \@firstoftwo
 \else \expandafter \@secondoftwo
 \fi
}%
\providecommand \natexlab [1]{#1}%
\providecommand \enquote  [1]{``#1''}%
\providecommand \bibnamefont  [1]{#1}%
\providecommand \bibfnamefont [1]{#1}%
\providecommand \citenamefont [1]{#1}%
\providecommand \href@noop [0]{\@secondoftwo}%
\providecommand \href [0]{\begingroup \@sanitize@url \@href}%
\providecommand \@href[1]{\@@startlink{#1}\@@href}%
\providecommand \@@href[1]{\endgroup#1\@@endlink}%
\providecommand \@sanitize@url [0]{\catcode `\\12\catcode `\$12\catcode
  `\&12\catcode `\#12\catcode `\^12\catcode `\_12\catcode `\%12\relax}%
\providecommand \@@startlink[1]{}%
\providecommand \@@endlink[0]{}%
\providecommand \url  [0]{\begingroup\@sanitize@url \@url }%
\providecommand \@url [1]{\endgroup\@href {#1}{\urlprefix }}%
\providecommand \urlprefix  [0]{URL }%
\providecommand \Eprint [0]{\href }%
\providecommand \doibase [0]{http://dx.doi.org/}%
\providecommand \selectlanguage [0]{\@gobble}%
\providecommand \bibinfo  [0]{\@secondoftwo}%
\providecommand \bibfield  [0]{\@secondoftwo}%
\providecommand \translation [1]{[#1]}%
\providecommand \BibitemOpen [0]{}%
\providecommand \bibitemStop [0]{}%
\providecommand \bibitemNoStop [0]{.\EOS\space}%
\providecommand \EOS [0]{\spacefactor3000\relax}%
\providecommand \BibitemShut  [1]{\csname bibitem#1\endcsname}%
\let\auto@bib@innerbib\@empty
\bibitem [{\citenamefont {Tippins}(1965)}]{tippins1965optical}%
  \BibitemOpen
  \bibfield  {author} {\bibinfo {author} {\bibfnamefont {H.}~\bibnamefont
  {Tippins}},\ }\href
  {https://journals.aps.org/pr/abstract/10.1103/PhysRev.140.A316} {\bibfield
  {journal} {\bibinfo  {journal} {Phys. Rev.}\ }\textbf {\bibinfo {volume}
  {140}},\ \bibinfo {pages} {A316} (\bibinfo {year} {1965})}\BibitemShut
  {NoStop}%
\bibitem [{\citenamefont {Matsumoto}\ \emph {et~al.}(1974)\citenamefont
  {Matsumoto}, \citenamefont {Aoki}, \citenamefont {Kinoshita},\ and\
  \citenamefont {Aono}}]{matsumoto1974absorption}%
  \BibitemOpen
  \bibfield  {author} {\bibinfo {author} {\bibfnamefont {T.}~\bibnamefont
  {Matsumoto}}, \bibinfo {author} {\bibfnamefont {M.}~\bibnamefont {Aoki}},
  \bibinfo {author} {\bibfnamefont {A.}~\bibnamefont {Kinoshita}}, \ and\
  \bibinfo {author} {\bibfnamefont {T.}~\bibnamefont {Aono}},\ }\href
  {https://journals.aps.org/pr/abstract/10.1103/PhysRev.140.A316} {\bibfield
  {journal} {\bibinfo  {journal} {Jpn. J. Appl. Phys}\ }\textbf {\bibinfo
  {volume} {13}},\ \bibinfo {pages} {1578} (\bibinfo {year}
  {1974})}\BibitemShut {NoStop}%
\bibitem [{\citenamefont {Sturm}\ \emph {et~al.}(2016)\citenamefont {Sturm},
  \citenamefont {Schmidt-Grund}, \citenamefont {Kranert}, \citenamefont
  {Furthm{\"u}ller}, \citenamefont {Bechstedt},\ and\ \citenamefont
  {Grundmann}}]{sturm2016dipole}%
  \BibitemOpen
  \bibfield  {author} {\bibinfo {author} {\bibfnamefont {C.}~\bibnamefont
  {Sturm}}, \bibinfo {author} {\bibfnamefont {R.}~\bibnamefont
  {Schmidt-Grund}}, \bibinfo {author} {\bibfnamefont {C.}~\bibnamefont
  {Kranert}}, \bibinfo {author} {\bibfnamefont {J.}~\bibnamefont
  {Furthm{\"u}ller}}, \bibinfo {author} {\bibfnamefont {F.}~\bibnamefont
  {Bechstedt}}, \ and\ \bibinfo {author} {\bibfnamefont {M.}~\bibnamefont
  {Grundmann}},\ }\href
  {https://journals.aps.org/prb/pdf/10.1103/PhysRevB.96.245205} {\bibfield
  {journal} {\bibinfo  {journal} {Phys. Rev. B}\ }\textbf {\bibinfo {volume}
  {94}},\ \bibinfo {pages} {035148} (\bibinfo {year} {2016})}\BibitemShut
  {NoStop}%
\bibitem [{\citenamefont {Mock}\ \emph {et~al.}(2017)\citenamefont {Mock},
  \citenamefont {Korlacki}, \citenamefont {Briley}, \citenamefont
  {Darakchieva}, \citenamefont {Monemar}, \citenamefont {Kumagai},
  \citenamefont {Goto}, \citenamefont {Higashiwaki},\ and\ \citenamefont
  {Schubert}}]{mock2017band}%
  \BibitemOpen
  \bibfield  {author} {\bibinfo {author} {\bibfnamefont {A.}~\bibnamefont
  {Mock}}, \bibinfo {author} {\bibfnamefont {R.}~\bibnamefont {Korlacki}},
  \bibinfo {author} {\bibfnamefont {C.}~\bibnamefont {Briley}}, \bibinfo
  {author} {\bibfnamefont {V.}~\bibnamefont {Darakchieva}}, \bibinfo {author}
  {\bibfnamefont {B.}~\bibnamefont {Monemar}}, \bibinfo {author} {\bibfnamefont
  {Y.}~\bibnamefont {Kumagai}}, \bibinfo {author} {\bibfnamefont
  {K.}~\bibnamefont {Goto}}, \bibinfo {author} {\bibfnamefont {M.}~\bibnamefont
  {Higashiwaki}}, \ and\ \bibinfo {author} {\bibfnamefont {M.}~\bibnamefont
  {Schubert}},\ }\href
  {https://journals.aps.org/prb/pdf/10.1103/PhysRevB.96.245205} {\bibfield
  {journal} {\bibinfo  {journal} {Phys. Rev. B}\ }\textbf {\bibinfo {volume}
  {96}},\ \bibinfo {pages} {245205} (\bibinfo {year} {2017})}\BibitemShut
  {NoStop}%
\bibitem [{\citenamefont {Higashiwaki}\ \emph {et~al.}(2012)\citenamefont
  {Higashiwaki}, \citenamefont {Sasaki}, \citenamefont {Kuramata},
  \citenamefont {Masui},\ and\ \citenamefont
  {Yamakoshi}}]{higashiwaki2012gallium}%
  \BibitemOpen
  \bibfield  {author} {\bibinfo {author} {\bibfnamefont {M.}~\bibnamefont
  {Higashiwaki}}, \bibinfo {author} {\bibfnamefont {K.}~\bibnamefont {Sasaki}},
  \bibinfo {author} {\bibfnamefont {A.}~\bibnamefont {Kuramata}}, \bibinfo
  {author} {\bibfnamefont {T.}~\bibnamefont {Masui}}, \ and\ \bibinfo {author}
  {\bibfnamefont {S.}~\bibnamefont {Yamakoshi}},\ }\href
  {https://aip.scitation.org/doi/full/10.1063/1.3674287} {\bibfield  {journal}
  {\bibinfo  {journal} {Appl. Phys. Lett.}\ }\textbf {\bibinfo {volume}
  {100}},\ \bibinfo {pages} {013504} (\bibinfo {year} {2012})}\BibitemShut
  {NoStop}%
\bibitem [{\citenamefont {Suzuki}\ \emph {et~al.}(2009)\citenamefont {Suzuki},
  \citenamefont {Nakagomi}, \citenamefont {Kokubun}, \citenamefont {Arai},\
  and\ \citenamefont {Ohira}}]{suzuki2009enhancement}%
  \BibitemOpen
  \bibfield  {author} {\bibinfo {author} {\bibfnamefont {R.}~\bibnamefont
  {Suzuki}}, \bibinfo {author} {\bibfnamefont {S.}~\bibnamefont {Nakagomi}},
  \bibinfo {author} {\bibfnamefont {Y.}~\bibnamefont {Kokubun}}, \bibinfo
  {author} {\bibfnamefont {N.}~\bibnamefont {Arai}}, \ and\ \bibinfo {author}
  {\bibfnamefont {S.}~\bibnamefont {Ohira}},\ }\href
  {https://aip.scitation.org/doi/10.1063/1.3147197} {\bibfield  {journal}
  {\bibinfo  {journal} {Appl. Phys. Lett.}\ }\textbf {\bibinfo {volume} {94}},\
  \bibinfo {pages} {222102} (\bibinfo {year} {2009})}\BibitemShut {NoStop}%
\bibitem [{\citenamefont {Oshima}\ \emph {et~al.}(2008)\citenamefont {Oshima},
  \citenamefont {Okuno}, \citenamefont {Arai}, \citenamefont {Suzuki},
  \citenamefont {Ohira},\ and\ \citenamefont {Fujita}}]{oshima2008vertical}%
  \BibitemOpen
  \bibfield  {author} {\bibinfo {author} {\bibfnamefont {T.}~\bibnamefont
  {Oshima}}, \bibinfo {author} {\bibfnamefont {T.}~\bibnamefont {Okuno}},
  \bibinfo {author} {\bibfnamefont {N.}~\bibnamefont {Arai}}, \bibinfo {author}
  {\bibfnamefont {N.}~\bibnamefont {Suzuki}}, \bibinfo {author} {\bibfnamefont
  {S.}~\bibnamefont {Ohira}}, \ and\ \bibinfo {author} {\bibfnamefont
  {S.}~\bibnamefont {Fujita}},\ }\href
  {https://iopscience.iop.org/article/10.1143/APEX.1.011202} {\bibfield
  {journal} {\bibinfo  {journal} {Appl. Phys. Express}\ }\textbf {\bibinfo
  {volume} {1}},\ \bibinfo {pages} {011202} (\bibinfo {year}
  {2008})}\BibitemShut {NoStop}%
\bibitem [{\citenamefont {Alema}\ \emph {et~al.}(2019)\citenamefont {Alema},
  \citenamefont {Hertog}, \citenamefont {Mukhopadhyay}, \citenamefont {Zhang},
  \citenamefont {Mauze}, \citenamefont {Osinsky}, \citenamefont {Schoenfeld},
  \citenamefont {Speck},\ and\ \citenamefont {Vogt}}]{alema2019solar}%
  \BibitemOpen
  \bibfield  {author} {\bibinfo {author} {\bibfnamefont {F.}~\bibnamefont
  {Alema}}, \bibinfo {author} {\bibfnamefont {B.}~\bibnamefont {Hertog}},
  \bibinfo {author} {\bibfnamefont {P.}~\bibnamefont {Mukhopadhyay}}, \bibinfo
  {author} {\bibfnamefont {Y.}~\bibnamefont {Zhang}}, \bibinfo {author}
  {\bibfnamefont {A.}~\bibnamefont {Mauze}}, \bibinfo {author} {\bibfnamefont
  {A.}~\bibnamefont {Osinsky}}, \bibinfo {author} {\bibfnamefont {W.~V.}\
  \bibnamefont {Schoenfeld}}, \bibinfo {author} {\bibfnamefont {J.~S.}\
  \bibnamefont {Speck}}, \ and\ \bibinfo {author} {\bibfnamefont
  {T.}~\bibnamefont {Vogt}},\ }\href
  {https://aip.scitation.org/doi/full/10.1063/1.5064471\%40apm.2019.WBO2019.issue-1}
  {\bibfield  {journal} {\bibinfo  {journal} {APL Materials}\ }\textbf
  {\bibinfo {volume} {7}},\ \bibinfo {pages} {022527} (\bibinfo {year}
  {2019})}\BibitemShut {NoStop}%
\bibitem [{\citenamefont {Feng}\ \emph {et~al.}(2019)\citenamefont {Feng},
  \citenamefont {{Anhar Uddin Bhuiyan}}, \citenamefont {Karim},\ and\
  \citenamefont {Zhao}}]{feng2019mocvd}%
  \BibitemOpen
  \bibfield  {author} {\bibinfo {author} {\bibfnamefont {Z.}~\bibnamefont
  {Feng}}, \bibinfo {author} {\bibfnamefont {A.~F.~M.}\ \bibnamefont {{Anhar
  Uddin Bhuiyan}}}, \bibinfo {author} {\bibfnamefont {M.~R.}\ \bibnamefont
  {Karim}}, \ and\ \bibinfo {author} {\bibfnamefont {H.}~\bibnamefont {Zhao}},\
  }\href {https://aip.scitation.org/doi/full/10.1063/1.5109678} {\bibfield
  {journal} {\bibinfo  {journal} {Appl. Phys. Lett.}\ }\textbf {\bibinfo
  {volume} {114}},\ \bibinfo {pages} {250601} (\bibinfo {year}
  {2019})}\BibitemShut {NoStop}%
\bibitem [{\citenamefont {Zhang}\ \emph {et~al.}(2019)\citenamefont {Zhang},
  \citenamefont {Alema}, \citenamefont {Mauze}, \citenamefont {Koksaldi},
  \citenamefont {Miller}, \citenamefont {Osinsky},\ and\ \citenamefont
  {Speck}}]{zhang2019mocvd}%
  \BibitemOpen
  \bibfield  {author} {\bibinfo {author} {\bibfnamefont {Y.}~\bibnamefont
  {Zhang}}, \bibinfo {author} {\bibfnamefont {F.}~\bibnamefont {Alema}},
  \bibinfo {author} {\bibfnamefont {A.}~\bibnamefont {Mauze}}, \bibinfo
  {author} {\bibfnamefont {O.~S.}\ \bibnamefont {Koksaldi}}, \bibinfo {author}
  {\bibfnamefont {R.}~\bibnamefont {Miller}}, \bibinfo {author} {\bibfnamefont
  {A.}~\bibnamefont {Osinsky}}, \ and\ \bibinfo {author} {\bibfnamefont
  {J.~S.}\ \bibnamefont {Speck}},\ }\href
  {https://aip.scitation.org/doi/full/10.1063/1.5058059} {\bibfield  {journal}
  {\bibinfo  {journal} {APL Materials}\ }\textbf {\bibinfo {volume} {7}},\
  \bibinfo {pages} {22506} (\bibinfo {year} {2019})}\BibitemShut {NoStop}%
\bibitem [{\citenamefont {Son}\ \emph {et~al.}(2016)\citenamefont {Son},
  \citenamefont {Goto}, \citenamefont {Nomura}, \citenamefont {Thieu},
  \citenamefont {Togashi}, \citenamefont {Murakami}, \citenamefont {Kumagai},
  \citenamefont {Kuramata}, \citenamefont {Higashiwaki}, \citenamefont
  {Koukitu}, \citenamefont {Yamakoshi}, \citenamefont {Monemar},\ and\
  \citenamefont {Janzen}}]{son2016electronic}%
  \BibitemOpen
  \bibfield  {author} {\bibinfo {author} {\bibfnamefont {N.}~\bibnamefont
  {Son}}, \bibinfo {author} {\bibfnamefont {K.}~\bibnamefont {Goto}}, \bibinfo
  {author} {\bibfnamefont {K.}~\bibnamefont {Nomura}}, \bibinfo {author}
  {\bibfnamefont {Q.}~\bibnamefont {Thieu}}, \bibinfo {author} {\bibfnamefont
  {R.}~\bibnamefont {Togashi}}, \bibinfo {author} {\bibfnamefont
  {H.}~\bibnamefont {Murakami}}, \bibinfo {author} {\bibfnamefont
  {Y.}~\bibnamefont {Kumagai}}, \bibinfo {author} {\bibfnamefont
  {A.}~\bibnamefont {Kuramata}}, \bibinfo {author} {\bibfnamefont
  {M.}~\bibnamefont {Higashiwaki}}, \bibinfo {author} {\bibfnamefont
  {A.}~\bibnamefont {Koukitu}}, \bibinfo {author} {\bibfnamefont
  {S.}~\bibnamefont {Yamakoshi}}, \bibinfo {author} {\bibfnamefont
  {B.}~\bibnamefont {Monemar}}, \ and\ \bibinfo {author} {\bibfnamefont
  {E.}~\bibnamefont {Janzen}},\ }\href
  {https://aip.scitation.org/doi/full/10.1063/1.4972040} {\bibfield  {journal}
  {\bibinfo  {journal} {J. Appl. Phys.}\ }\textbf {\bibinfo {volume} {120}},\
  \bibinfo {pages} {235703} (\bibinfo {year} {2016})}\BibitemShut {NoStop}%
\bibitem [{\citenamefont {Parisini}\ and\ \citenamefont
  {Fornari}(2016)}]{parisini2016analysis}%
  \BibitemOpen
  \bibfield  {author} {\bibinfo {author} {\bibfnamefont {A.}~\bibnamefont
  {Parisini}}\ and\ \bibinfo {author} {\bibfnamefont {R.}~\bibnamefont
  {Fornari}},\ }\href
  {https://iopscience.iop.org/article/10.1088/0268-1242/31/3/035023/meta}
  {\bibfield  {journal} {\bibinfo  {journal} {Semicond. Sci. Technol.}\
  }\textbf {\bibinfo {volume} {31}},\ \bibinfo {pages} {035023} (\bibinfo
  {year} {2016})}\BibitemShut {NoStop}%
\bibitem [{\citenamefont {Ma}\ \emph {et~al.}(2016)\citenamefont {Ma},
  \citenamefont {Tanen}, \citenamefont {Verma}, \citenamefont {Guo},
  \citenamefont {Luo}, \citenamefont {Xing},\ and\ \citenamefont
  {Jena}}]{ma2016intrinsic}%
  \BibitemOpen
  \bibfield  {author} {\bibinfo {author} {\bibfnamefont {N.}~\bibnamefont
  {Ma}}, \bibinfo {author} {\bibfnamefont {N.}~\bibnamefont {Tanen}}, \bibinfo
  {author} {\bibfnamefont {A.}~\bibnamefont {Verma}}, \bibinfo {author}
  {\bibfnamefont {Z.}~\bibnamefont {Guo}}, \bibinfo {author} {\bibfnamefont
  {T.}~\bibnamefont {Luo}}, \bibinfo {author} {\bibfnamefont {H.}~\bibnamefont
  {Xing}}, \ and\ \bibinfo {author} {\bibfnamefont {D.}~\bibnamefont {Jena}},\
  }\href
  {https://aip.scitation.org/doi/full/10.1063/1.4968550\%40apl.2018.GAO2018.issue-1}
  {\bibfield  {journal} {\bibinfo  {journal} {Appl. Phys. Lett.}\ }\textbf
  {\bibinfo {volume} {109}},\ \bibinfo {pages} {212101} (\bibinfo {year}
  {2016})}\BibitemShut {NoStop}%
\bibitem [{\citenamefont {Higashiwaki}\ \emph {et~al.}(2017)\citenamefont
  {Higashiwaki}, \citenamefont {Kuramata}, \citenamefont {Murakami},\ and\
  \citenamefont {Kumagai}}]{higashiwaki2017state}%
  \BibitemOpen
  \bibfield  {author} {\bibinfo {author} {\bibfnamefont {M.}~\bibnamefont
  {Higashiwaki}}, \bibinfo {author} {\bibfnamefont {A.}~\bibnamefont
  {Kuramata}}, \bibinfo {author} {\bibfnamefont {H.}~\bibnamefont {Murakami}},
  \ and\ \bibinfo {author} {\bibfnamefont {Y.}~\bibnamefont {Kumagai}},\ }\href
  {https://iopscience.iop.org/article/10.1088/1361-6463/aa7aff/meta} {\bibfield
   {journal} {\bibinfo  {journal} {J. Phys. D}\ }\textbf {\bibinfo {volume}
  {50}},\ \bibinfo {pages} {333002} (\bibinfo {year} {2017})}\BibitemShut
  {NoStop}%
\bibitem [{\citenamefont {Moser}\ \emph {et~al.}(2017)\citenamefont {Moser},
  \citenamefont {McCandless}, \citenamefont {Crespo}, \citenamefont {Leedy},
  \citenamefont {Green}, \citenamefont {Neal}, \citenamefont {Mou},
  \citenamefont {Ahmadi}, \citenamefont {Speck}, \citenamefont {Chabak},
  \citenamefont {Peixoto},\ and\ \citenamefont {Jesse}}]{moser2017ge}%
  \BibitemOpen
  \bibfield  {author} {\bibinfo {author} {\bibfnamefont {N.}~\bibnamefont
  {Moser}}, \bibinfo {author} {\bibfnamefont {J.}~\bibnamefont {McCandless}},
  \bibinfo {author} {\bibfnamefont {A.}~\bibnamefont {Crespo}}, \bibinfo
  {author} {\bibfnamefont {K.}~\bibnamefont {Leedy}}, \bibinfo {author}
  {\bibfnamefont {A.}~\bibnamefont {Green}}, \bibinfo {author} {\bibfnamefont
  {A.}~\bibnamefont {Neal}}, \bibinfo {author} {\bibfnamefont {S.}~\bibnamefont
  {Mou}}, \bibinfo {author} {\bibfnamefont {E.}~\bibnamefont {Ahmadi}},
  \bibinfo {author} {\bibfnamefont {J.}~\bibnamefont {Speck}}, \bibinfo
  {author} {\bibfnamefont {K.}~\bibnamefont {Chabak}}, \bibinfo {author}
  {\bibfnamefont {N.}~\bibnamefont {Peixoto}}, \ and\ \bibinfo {author}
  {\bibfnamefont {G.}~\bibnamefont {Jesse}},\ }\href
  {https://ieeexplore.ieee.org/abstract/document/7911200} {\bibfield  {journal}
  {\bibinfo  {journal} {IEEE Electron Device Lett.}\ }\textbf {\bibinfo
  {volume} {38}},\ \bibinfo {pages} {775} (\bibinfo {year} {2017})}\BibitemShut
  {NoStop}%
\bibitem [{\citenamefont {Neal}\ \emph {et~al.}(2018)\citenamefont {Neal},
  \citenamefont {Mou}, \citenamefont {Rafique}, \citenamefont {Zhao},
  \citenamefont {Ahmadi}, \citenamefont {Speck}, \citenamefont {Stevens},
  \citenamefont {Blevins}, \citenamefont {Thomson}, \citenamefont {Moser},\
  and\ \citenamefont {Others}}]{neal2018donors}%
  \BibitemOpen
  \bibfield  {author} {\bibinfo {author} {\bibfnamefont {A.~T.}\ \bibnamefont
  {Neal}}, \bibinfo {author} {\bibfnamefont {S.}~\bibnamefont {Mou}}, \bibinfo
  {author} {\bibfnamefont {S.}~\bibnamefont {Rafique}}, \bibinfo {author}
  {\bibfnamefont {H.}~\bibnamefont {Zhao}}, \bibinfo {author} {\bibfnamefont
  {E.}~\bibnamefont {Ahmadi}}, \bibinfo {author} {\bibfnamefont {J.~S.}\
  \bibnamefont {Speck}}, \bibinfo {author} {\bibfnamefont {K.~T.}\ \bibnamefont
  {Stevens}}, \bibinfo {author} {\bibfnamefont {J.~D.}\ \bibnamefont
  {Blevins}}, \bibinfo {author} {\bibfnamefont {D.~B.}\ \bibnamefont
  {Thomson}}, \bibinfo {author} {\bibfnamefont {N.}~\bibnamefont {Moser}}, \
  and\ \bibinfo {author} {\bibnamefont {Others}},\ }\href
  {https://aip.scitation.org/doi/full/10.1063/1.5034474} {\bibfield  {journal}
  {\bibinfo  {journal} {Appl. Phys. Lett.}\ }\textbf {\bibinfo {volume}
  {113}},\ \bibinfo {pages} {62101} (\bibinfo {year} {2018})}\BibitemShut
  {NoStop}%
\bibitem [{\citenamefont {Orita}\ \emph {et~al.}(2000)\citenamefont {Orita},
  \citenamefont {Ohta}, \citenamefont {Hirano},\ and\ \citenamefont
  {Hosono}}]{orita2000deep}%
  \BibitemOpen
  \bibfield  {author} {\bibinfo {author} {\bibfnamefont {M.}~\bibnamefont
  {Orita}}, \bibinfo {author} {\bibfnamefont {H.}~\bibnamefont {Ohta}},
  \bibinfo {author} {\bibfnamefont {M.}~\bibnamefont {Hirano}}, \ and\ \bibinfo
  {author} {\bibfnamefont {H.}~\bibnamefont {Hosono}},\ }\href
  {https://aip.scitation.org/doi/abs/10.1063/1.1330559} {\bibfield  {journal}
  {\bibinfo  {journal} {Appl. Phys. Lett.}\ }\textbf {\bibinfo {volume} {77}},\
  \bibinfo {pages} {4166} (\bibinfo {year} {2000})}\BibitemShut {NoStop}%
\bibitem [{\citenamefont {Oishi}\ \emph {et~al.}(2016)\citenamefont {Oishi},
  \citenamefont {Harada}, \citenamefont {Koga},\ and\ \citenamefont
  {Kasu}}]{oishi2016conduction}%
  \BibitemOpen
  \bibfield  {author} {\bibinfo {author} {\bibfnamefont {T.}~\bibnamefont
  {Oishi}}, \bibinfo {author} {\bibfnamefont {K.}~\bibnamefont {Harada}},
  \bibinfo {author} {\bibfnamefont {Y.}~\bibnamefont {Koga}}, \ and\ \bibinfo
  {author} {\bibfnamefont {M.}~\bibnamefont {Kasu}},\ }\href
  {https://iopscience.iop.org/article/10.7567/JJAP.55.030305/meta} {\bibfield
  {journal} {\bibinfo  {journal} {Jpn. J. Appl. Phys.}\ }\textbf {\bibinfo
  {volume} {55}},\ \bibinfo {pages} {30305} (\bibinfo {year}
  {2016})}\BibitemShut {NoStop}%
\bibitem [{\citenamefont {Higashiwaki}\ \emph {et~al.}(2013)\citenamefont
  {Higashiwaki}, \citenamefont {Sasaki}, \citenamefont {Kamimura},
  \citenamefont {Hoi~Wong}, \citenamefont {Krishnamurthy}, \citenamefont
  {Kuramata}, \citenamefont {Masui},\ and\ \citenamefont
  {Yamakoshi}}]{higashiwaki2013depletion}%
  \BibitemOpen
  \bibfield  {author} {\bibinfo {author} {\bibfnamefont {M.}~\bibnamefont
  {Higashiwaki}}, \bibinfo {author} {\bibfnamefont {K.}~\bibnamefont {Sasaki}},
  \bibinfo {author} {\bibfnamefont {T.}~\bibnamefont {Kamimura}}, \bibinfo
  {author} {\bibfnamefont {M.}~\bibnamefont {Hoi~Wong}}, \bibinfo {author}
  {\bibfnamefont {D.}~\bibnamefont {Krishnamurthy}}, \bibinfo {author}
  {\bibfnamefont {A.}~\bibnamefont {Kuramata}}, \bibinfo {author}
  {\bibfnamefont {T.}~\bibnamefont {Masui}}, \ and\ \bibinfo {author}
  {\bibfnamefont {S.}~\bibnamefont {Yamakoshi}},\ }\href
  {https://aip.scitation.org/doi/full/10.1063/1.4821858\%40apl.2018.GAO2018.issue-1}
  {\bibfield  {journal} {\bibinfo  {journal} {Appl. Phys. Lett.}\ }\textbf
  {\bibinfo {volume} {103}},\ \bibinfo {pages} {123511} (\bibinfo {year}
  {2013})}\BibitemShut {NoStop}%
\bibitem [{\citenamefont {Varley}\ \emph {et~al.}(2010)\citenamefont {Varley},
  \citenamefont {Weber}, \citenamefont {Janotti},\ and\ \citenamefont {Van~de
  Walle}}]{varley2010oxygen}%
  \BibitemOpen
  \bibfield  {author} {\bibinfo {author} {\bibfnamefont {J.~B.}\ \bibnamefont
  {Varley}}, \bibinfo {author} {\bibfnamefont {J.~R.}\ \bibnamefont {Weber}},
  \bibinfo {author} {\bibfnamefont {A.}~\bibnamefont {Janotti}}, \ and\
  \bibinfo {author} {\bibfnamefont {C.~G.}\ \bibnamefont {Van~de Walle}},\
  }\href
  {https://aip.scitation.org/doi/full/10.1063/1.3499306\%40apl.2018.GAO2018.issue-1}
  {\bibfield  {journal} {\bibinfo  {journal} {Appl. Phys. Lett.}\ }\textbf
  {\bibinfo {volume} {97}},\ \bibinfo {pages} {142106} (\bibinfo {year}
  {2010})}\BibitemShut {NoStop}%
\bibitem [{\citenamefont {Varley}\ \emph {et~al.}(2020)\citenamefont {Varley},
  \citenamefont {Perron}, \citenamefont {Lordi}, \citenamefont
  {Wickramaratne},\ and\ \citenamefont {Lyons}}]{varley2020prospects}%
  \BibitemOpen
  \bibfield  {author} {\bibinfo {author} {\bibfnamefont {J.~B.}\ \bibnamefont
  {Varley}}, \bibinfo {author} {\bibfnamefont {A.}~\bibnamefont {Perron}},
  \bibinfo {author} {\bibfnamefont {V.}~\bibnamefont {Lordi}}, \bibinfo
  {author} {\bibfnamefont {D.}~\bibnamefont {Wickramaratne}}, \ and\ \bibinfo
  {author} {\bibfnamefont {J.~L.}\ \bibnamefont {Lyons}},\ }\href
  {https://aip.scitation.org/doi/full/10.1063/5.0006224} {\bibfield  {journal}
  {\bibinfo  {journal} {Appl. Phys. Lett.}\ }\textbf {\bibinfo {volume}
  {116}},\ \bibinfo {pages} {172104} (\bibinfo {year} {2020})}\BibitemShut
  {NoStop}%
\bibitem [{\citenamefont {Lyons}\ \emph
  {et~al.}(2014{\natexlab{a}})\citenamefont {Lyons}, \citenamefont {Steiauf},
  \citenamefont {Janotti},\ and\ \citenamefont {Van~de
  Walle}}]{lyons2014carbon}%
  \BibitemOpen
  \bibfield  {author} {\bibinfo {author} {\bibfnamefont {J.~L.}\ \bibnamefont
  {Lyons}}, \bibinfo {author} {\bibfnamefont {D.}~\bibnamefont {Steiauf}},
  \bibinfo {author} {\bibfnamefont {A.}~\bibnamefont {Janotti}}, \ and\
  \bibinfo {author} {\bibfnamefont {C.~G.}\ \bibnamefont {Van~de Walle}},\
  }\href
  {https://journals.aps.org/prapplied/abstract/10.1103/PhysRevApplied.2.064005}
  {\bibfield  {journal} {\bibinfo  {journal} {Phys. Rev. Appl.}\ }\textbf
  {\bibinfo {volume} {2}},\ \bibinfo {pages} {064005} (\bibinfo {year}
  {2014}{\natexlab{a}})}\BibitemShut {NoStop}%
\bibitem [{\citenamefont {Varley}\ \emph {et~al.}(2011)\citenamefont {Varley},
  \citenamefont {Peelaers}, \citenamefont {Janotti},\ and\ \citenamefont
  {Van~de Walle}}]{varley2011hydrogenated}%
  \BibitemOpen
  \bibfield  {author} {\bibinfo {author} {\bibfnamefont {J.~B.}\ \bibnamefont
  {Varley}}, \bibinfo {author} {\bibfnamefont {H.}~\bibnamefont {Peelaers}},
  \bibinfo {author} {\bibfnamefont {A.}~\bibnamefont {Janotti}}, \ and\
  \bibinfo {author} {\bibfnamefont {C.~G.}\ \bibnamefont {Van~de Walle}},\
  }\href
  {https://iopscience.iop.org/article/10.1088/0953-8984/23/33/334212/meta}
  {\bibfield  {journal} {\bibinfo  {journal} {J. Phys. Conden. Matter}\
  }\textbf {\bibinfo {volume} {23}},\ \bibinfo {pages} {334212} (\bibinfo
  {year} {2011})}\BibitemShut {NoStop}%
\bibitem [{\citenamefont {Ingebrigtsen}\ \emph {et~al.}(2019)\citenamefont
  {Ingebrigtsen}, \citenamefont {Kuznetsov}, \citenamefont {Svensson},
  \citenamefont {Alfieri}, \citenamefont {Mihaila}, \citenamefont
  {Badst{\"u}bner}, \citenamefont {Perron}, \citenamefont {Vines},\ and\
  \citenamefont {Varley}}]{ingebrigtsen2019impact}%
  \BibitemOpen
  \bibfield  {author} {\bibinfo {author} {\bibfnamefont {M.~E.}\ \bibnamefont
  {Ingebrigtsen}}, \bibinfo {author} {\bibfnamefont {A.~Y.}\ \bibnamefont
  {Kuznetsov}}, \bibinfo {author} {\bibfnamefont {B.~G.}\ \bibnamefont
  {Svensson}}, \bibinfo {author} {\bibfnamefont {G.}~\bibnamefont {Alfieri}},
  \bibinfo {author} {\bibfnamefont {A.}~\bibnamefont {Mihaila}}, \bibinfo
  {author} {\bibfnamefont {U.}~\bibnamefont {Badst{\"u}bner}}, \bibinfo
  {author} {\bibfnamefont {A.}~\bibnamefont {Perron}}, \bibinfo {author}
  {\bibfnamefont {L.}~\bibnamefont {Vines}}, \ and\ \bibinfo {author}
  {\bibfnamefont {J.~B.}\ \bibnamefont {Varley}},\ }\href
  {https://aip.scitation.org/doi/pdf/10.1063/1.5054826} {\bibfield  {journal}
  {\bibinfo  {journal} {APL Mater.}\ }\textbf {\bibinfo {volume} {7}},\
  \bibinfo {pages} {022510} (\bibinfo {year} {2019})}\BibitemShut {NoStop}%
\bibitem [{\citenamefont {von Bardeleben}\ and\ \citenamefont
  {Cantin}(2020)}]{von2020unusual}%
  \BibitemOpen
  \bibfield  {author} {\bibinfo {author} {\bibfnamefont {H.}~\bibnamefont {von
  Bardeleben}}\ and\ \bibinfo {author} {\bibfnamefont {J.}~\bibnamefont
  {Cantin}},\ }\href {https://aip.scitation.org/doi/abs/10.1063/5.0023546}
  {\bibfield  {journal} {\bibinfo  {journal} {J. Appl. Phys.}\ }\textbf
  {\bibinfo {volume} {128}},\ \bibinfo {pages} {125702} (\bibinfo {year}
  {2020})}\BibitemShut {NoStop}%
\bibitem [{\citenamefont {Chadi}\ and\ \citenamefont
  {Chang}(1988)}]{chadi1988theory}%
  \BibitemOpen
  \bibfield  {author} {\bibinfo {author} {\bibfnamefont {D.}~\bibnamefont
  {Chadi}}\ and\ \bibinfo {author} {\bibfnamefont {K.-J.}\ \bibnamefont
  {Chang}},\ }\href
  {https://journals.aps.org/prl/abstract/10.1103/PhysRevLett.61.873} {\bibfield
   {journal} {\bibinfo  {journal} {Phys. Rev. Lett.}\ }\textbf {\bibinfo
  {volume} {61}},\ \bibinfo {pages} {873} (\bibinfo {year} {1988})}\BibitemShut
  {NoStop}%
\bibitem [{\citenamefont {Gordon}\ \emph {et~al.}(2014)\citenamefont {Gordon},
  \citenamefont {Lyons}, \citenamefont {Janotti},\ and\ \citenamefont {Van~de
  Walle}}]{gordon2014hybrid}%
  \BibitemOpen
  \bibfield  {author} {\bibinfo {author} {\bibfnamefont {L.}~\bibnamefont
  {Gordon}}, \bibinfo {author} {\bibfnamefont {J.~L.}\ \bibnamefont {Lyons}},
  \bibinfo {author} {\bibfnamefont {A.}~\bibnamefont {Janotti}}, \ and\
  \bibinfo {author} {\bibfnamefont {C.~G.}\ \bibnamefont {Van~de Walle}},\
  }\href {https://journals.aps.org/prb/abstract/10.1103/PhysRevB.89.085204}
  {\bibfield  {journal} {\bibinfo  {journal} {Phys. Rev. B}\ }\textbf {\bibinfo
  {volume} {89}},\ \bibinfo {pages} {085204} (\bibinfo {year}
  {2014})}\BibitemShut {NoStop}%
\bibitem [{\citenamefont {Peelaers}\ \emph {et~al.}(2018)\citenamefont
  {Peelaers}, \citenamefont {Varley}, \citenamefont {Speck},\ and\
  \citenamefont {Van~de Walle}}]{peelaers2018structural}%
  \BibitemOpen
  \bibfield  {author} {\bibinfo {author} {\bibfnamefont {H.}~\bibnamefont
  {Peelaers}}, \bibinfo {author} {\bibfnamefont {J.~B.}\ \bibnamefont
  {Varley}}, \bibinfo {author} {\bibfnamefont {J.~S.}\ \bibnamefont {Speck}}, \
  and\ \bibinfo {author} {\bibfnamefont {C.~G.}\ \bibnamefont {Van~de Walle}},\
  }\href {https://aip.scitation.org/doi/full/10.1063/1.5036991} {\bibfield
  {journal} {\bibinfo  {journal} {Appl. Phys. Lett.}\ }\textbf {\bibinfo
  {volume} {112}},\ \bibinfo {pages} {242101} (\bibinfo {year}
  {2018})}\BibitemShut {NoStop}%
\bibitem [{\citenamefont {Peelaers}\ \emph
  {et~al.}(2019{\natexlab{a}})\citenamefont {Peelaers}, \citenamefont {Varley},
  \citenamefont {Speck},\ and\ \citenamefont {Van~de
  Walle}}]{peelaers2019erratum}%
  \BibitemOpen
  \bibfield  {author} {\bibinfo {author} {\bibfnamefont {H.}~\bibnamefont
  {Peelaers}}, \bibinfo {author} {\bibfnamefont {J.~B.}\ \bibnamefont
  {Varley}}, \bibinfo {author} {\bibfnamefont {J.~S.}\ \bibnamefont {Speck}}, \
  and\ \bibinfo {author} {\bibfnamefont {C.~G.}\ \bibnamefont {Van~de Walle}},\
  }\href {https://aip.scitation.org/doi/full/10.1063/1.5127763} {\bibfield
  {journal} {\bibinfo  {journal} {Appl. Phys. Lett.}\ }\textbf {\bibinfo
  {volume} {115}},\ \bibinfo {pages} {159901} (\bibinfo {year}
  {2019}{\natexlab{a}})},\ \bibinfo {note} {(erratum)}\BibitemShut {NoStop}%
\bibitem [{\citenamefont {Varley}(2021)}]{Varley2021_review}%
  \BibitemOpen
  \bibfield  {author} {\bibinfo {author} {\bibfnamefont {J.~B.}\ \bibnamefont
  {Varley}},\ }\href {\doibase 10.1557/s43578-021-00371-7} {\bibfield
  {journal} {\bibinfo  {journal} {J. Mater. Res.}\ ,\ \bibinfo {pages} {1}}
  (\bibinfo {year} {2021})}\BibitemShut {NoStop}%
\bibitem [{\citenamefont {Zhang}\ \emph {et~al.}(2018)\citenamefont {Zhang},
  \citenamefont {Joishi}, \citenamefont {Xia}, \citenamefont {Brenner},
  \citenamefont {Lodha},\ and\ \citenamefont {Rajan}}]{zhang2018demonstration}%
  \BibitemOpen
  \bibfield  {author} {\bibinfo {author} {\bibfnamefont {Y.}~\bibnamefont
  {Zhang}}, \bibinfo {author} {\bibfnamefont {C.}~\bibnamefont {Joishi}},
  \bibinfo {author} {\bibfnamefont {Z.}~\bibnamefont {Xia}}, \bibinfo {author}
  {\bibfnamefont {M.}~\bibnamefont {Brenner}}, \bibinfo {author} {\bibfnamefont
  {S.}~\bibnamefont {Lodha}}, \ and\ \bibinfo {author} {\bibfnamefont
  {S.}~\bibnamefont {Rajan}},\ }\href
  {https://aip.scitation.org/doi/pdf/10.1063/1.5037095} {\bibfield  {journal}
  {\bibinfo  {journal} {Appl. Phys. Lett.}\ }\textbf {\bibinfo {volume}
  {112}},\ \bibinfo {pages} {233503} (\bibinfo {year} {2018})}\BibitemShut
  {NoStop}%
\bibitem [{\citenamefont {Zhao}()}]{zhao2020}%
  \BibitemOpen
  \bibfield  {author} {\bibinfo {author} {\bibfnamefont {H.~P.}\ \bibnamefont
  {Zhao}},\ }\href@noop {} {\emph {\bibinfo {title} {private
  communication}}}\BibitemShut {NoStop}%
\bibitem [{\citenamefont {Alema}\ \emph
  {et~al.}(2020{\natexlab{a}})\citenamefont {Alema}, \citenamefont {Zhang},
  \citenamefont {Osinsky}, \citenamefont {Orishchin}, \citenamefont {Valente},
  \citenamefont {Mauze},\ and\ \citenamefont {Speck}}]{alema2020}%
  \BibitemOpen
  \bibfield  {author} {\bibinfo {author} {\bibfnamefont {F.}~\bibnamefont
  {Alema}}, \bibinfo {author} {\bibfnamefont {Y.}~\bibnamefont {Zhang}},
  \bibinfo {author} {\bibfnamefont {A.}~\bibnamefont {Osinsky}}, \bibinfo
  {author} {\bibfnamefont {N.}~\bibnamefont {Orishchin}}, \bibinfo {author}
  {\bibfnamefont {N.}~\bibnamefont {Valente}}, \bibinfo {author} {\bibfnamefont
  {A.}~\bibnamefont {Mauze}}, \ and\ \bibinfo {author} {\bibfnamefont {J.~S.}\
  \bibnamefont {Speck}},\ }\href {\doibase 10.1063/1.5132752} {\bibfield
  {journal} {\bibinfo  {journal} {APL Materials}\ }\textbf {\bibinfo {volume}
  {8}},\ \bibinfo {pages} {021110} (\bibinfo {year}
  {2020}{\natexlab{a}})}\BibitemShut {NoStop}%
\bibitem [{\citenamefont {Seryogin}\ \emph {et~al.}(2020)\citenamefont
  {Seryogin}, \citenamefont {Alema}, \citenamefont {Valente}, \citenamefont
  {Fu}, \citenamefont {Steinbrunner}, \citenamefont {Neal}, \citenamefont
  {Mou}, \citenamefont {Fine},\ and\ \citenamefont
  {Osinsky}}]{seryogin2020mocvd}%
  \BibitemOpen
  \bibfield  {author} {\bibinfo {author} {\bibfnamefont {G.}~\bibnamefont
  {Seryogin}}, \bibinfo {author} {\bibfnamefont {F.}~\bibnamefont {Alema}},
  \bibinfo {author} {\bibfnamefont {N.}~\bibnamefont {Valente}}, \bibinfo
  {author} {\bibfnamefont {H.}~\bibnamefont {Fu}}, \bibinfo {author}
  {\bibfnamefont {E.}~\bibnamefont {Steinbrunner}}, \bibinfo {author}
  {\bibfnamefont {A.~T.}\ \bibnamefont {Neal}}, \bibinfo {author}
  {\bibfnamefont {S.}~\bibnamefont {Mou}}, \bibinfo {author} {\bibfnamefont
  {A.}~\bibnamefont {Fine}}, \ and\ \bibinfo {author} {\bibfnamefont
  {A.}~\bibnamefont {Osinsky}},\ }\href
  {https://aip.scitation.org/doi/full/10.1063/5.0031484} {\bibfield  {journal}
  {\bibinfo  {journal} {Appl. Phys. Lett.}\ }\textbf {\bibinfo {volume}
  {117}},\ \bibinfo {pages} {262101} (\bibinfo {year} {2020})}\BibitemShut
  {NoStop}%
\bibitem [{\citenamefont {Alema}\ \emph
  {et~al.}(2020{\natexlab{b}})\citenamefont {Alema}, \citenamefont {Zhang},
  \citenamefont {Mauze}, \citenamefont {Itoh}, \citenamefont {Speck},
  \citenamefont {Hertog},\ and\ \citenamefont {Osinsky}}]{alema2020h2o}%
  \BibitemOpen
  \bibfield  {author} {\bibinfo {author} {\bibfnamefont {F.}~\bibnamefont
  {Alema}}, \bibinfo {author} {\bibfnamefont {Y.}~\bibnamefont {Zhang}},
  \bibinfo {author} {\bibfnamefont {A.}~\bibnamefont {Mauze}}, \bibinfo
  {author} {\bibfnamefont {T.}~\bibnamefont {Itoh}}, \bibinfo {author}
  {\bibfnamefont {J.~S.}\ \bibnamefont {Speck}}, \bibinfo {author}
  {\bibfnamefont {B.}~\bibnamefont {Hertog}}, \ and\ \bibinfo {author}
  {\bibfnamefont {A.}~\bibnamefont {Osinsky}},\ }\href
  {https://aip.scitation.org/doi/abs/10.1063/5.0011910} {\bibfield  {journal}
  {\bibinfo  {journal} {AIP Advances}\ }\textbf {\bibinfo {volume} {10}},\
  \bibinfo {pages} {085002} (\bibinfo {year} {2020}{\natexlab{b}})}\BibitemShut
  {NoStop}%
\bibitem [{\citenamefont {Bl{\"{o}}chl}(1994)}]{Blochl1994}%
  \BibitemOpen
  \bibfield  {author} {\bibinfo {author} {\bibfnamefont {P.~E.}\ \bibnamefont
  {Bl{\"{o}}chl}},\ }\href {\doibase 10.1103/PhysRevB.50.17953} {\bibfield
  {journal} {\bibinfo  {journal} {Phys. Rev. B}\ }\textbf {\bibinfo {volume}
  {50}},\ \bibinfo {pages} {17953} (\bibinfo {year} {1994})}\BibitemShut
  {NoStop}%
\bibitem [{\citenamefont {Kresse}\ and\ \citenamefont
  {Hafner}(1993)}]{Kresse1993}%
  \BibitemOpen
  \bibfield  {author} {\bibinfo {author} {\bibfnamefont {G.}~\bibnamefont
  {Kresse}}\ and\ \bibinfo {author} {\bibfnamefont {J.}~\bibnamefont
  {Hafner}},\ }\href {\doibase 10.1103/PhysRevB.48.13115} {\bibfield  {journal}
  {\bibinfo  {journal} {Phys. Rev. B}\ }\textbf {\bibinfo {volume} {48}},\
  \bibinfo {pages} {13115} (\bibinfo {year} {1993})}\BibitemShut {NoStop}%
\bibitem [{\citenamefont {Kresse}\ and\ \citenamefont
  {Furthm{\"{u}}ller}(1996)}]{Kresse1996}%
  \BibitemOpen
  \bibfield  {author} {\bibinfo {author} {\bibfnamefont {G.}~\bibnamefont
  {Kresse}}\ and\ \bibinfo {author} {\bibfnamefont {J.}~\bibnamefont
  {Furthm{\"{u}}ller}},\ }\href
  {https://journals.aps.org/prb/abstract/10.1103/PhysRevB.54.11169} {\bibfield
  {journal} {\bibinfo  {journal} {Phys. Rev. B}\ }\textbf {\bibinfo {volume}
  {54}},\ \bibinfo {pages} {11169} (\bibinfo {year} {1996})}\BibitemShut
  {NoStop}%
\bibitem [{\citenamefont {Momma}\ and\ \citenamefont
  {Izumi}(2011)}]{momma2011vesta}%
  \BibitemOpen
  \bibfield  {author} {\bibinfo {author} {\bibfnamefont {K.}~\bibnamefont
  {Momma}}\ and\ \bibinfo {author} {\bibfnamefont {F.}~\bibnamefont {Izumi}},\
  }\href {https://scripts.iucr.org/cgi-bin/paper?db5098} {\bibfield  {journal}
  {\bibinfo  {journal} {J. Appl. Crystallogr.}\ }\textbf {\bibinfo {volume}
  {44}},\ \bibinfo {pages} {1272} (\bibinfo {year} {2011})}\BibitemShut
  {NoStop}%
\bibitem [{\citenamefont {Heyd}\ \emph {et~al.}(2003)\citenamefont {Heyd},
  \citenamefont {Scuseria},\ and\ \citenamefont {Ernzerhof}}]{heyd2003hybrid}%
  \BibitemOpen
  \bibfield  {author} {\bibinfo {author} {\bibfnamefont {J.}~\bibnamefont
  {Heyd}}, \bibinfo {author} {\bibfnamefont {G.~E.}\ \bibnamefont {Scuseria}},
  \ and\ \bibinfo {author} {\bibfnamefont {M.}~\bibnamefont {Ernzerhof}},\
  }\href {https://aip.scitation.org/doi/abs/10.1063/1.15640605} {\bibfield
  {journal} {\bibinfo  {journal} {J. Chem. Phys.}\ }\textbf {\bibinfo {volume}
  {118}},\ \bibinfo {pages} {8207} (\bibinfo {year} {2003})}\BibitemShut
  {NoStop}%
\bibitem [{\citenamefont {Heyd}\ and\ \citenamefont
  {Scuseria}(2006)}]{heyd2006erratum}%
  \BibitemOpen
  \bibfield  {author} {\bibinfo {author} {\bibfnamefont {J.}~\bibnamefont
  {Heyd}}\ and\ \bibinfo {author} {\bibfnamefont {G.~E.}\ \bibnamefont
  {Scuseria}},\ }\href {https://aip.scitation.org/doi/10.1063/1.2204597}
  {\bibfield  {journal} {\bibinfo  {journal} {J. Chem. Phys.}\ }\textbf
  {\bibinfo {volume} {124}},\ \bibinfo {pages} {219906} (\bibinfo {year}
  {2006})}\BibitemShut {NoStop}%
\bibitem [{\citenamefont {Franchy}\ \emph {et~al.}(1997)\citenamefont
  {Franchy}, \citenamefont {Schmitz}, \citenamefont {Gassmann},\ and\
  \citenamefont {Bartolucci}}]{franchy1997growth}%
  \BibitemOpen
  \bibfield  {author} {\bibinfo {author} {\bibfnamefont {R.}~\bibnamefont
  {Franchy}}, \bibinfo {author} {\bibfnamefont {G.}~\bibnamefont {Schmitz}},
  \bibinfo {author} {\bibfnamefont {P.}~\bibnamefont {Gassmann}}, \ and\
  \bibinfo {author} {\bibfnamefont {F.}~\bibnamefont {Bartolucci}},\ }\href
  {https://link.springer.com/article/10.1007/s003390050622} {\bibfield
  {journal} {\bibinfo  {journal} {Appl. Phys. A}\ }\textbf {\bibinfo {volume}
  {65}},\ \bibinfo {pages} {551} (\bibinfo {year} {1997})}\BibitemShut
  {NoStop}%
\bibitem [{\citenamefont {Kranert}\ \emph {et~al.}(2015)\citenamefont
  {Kranert}, \citenamefont {Jenderka}, \citenamefont {Lenzner}, \citenamefont
  {Lorenz}, \citenamefont {Von~Wenckstern}, \citenamefont {Schmidt-Grund},\
  and\ \citenamefont {Grundmann}}]{kranert2015lattice}%
  \BibitemOpen
  \bibfield  {author} {\bibinfo {author} {\bibfnamefont {C.}~\bibnamefont
  {Kranert}}, \bibinfo {author} {\bibfnamefont {M.}~\bibnamefont {Jenderka}},
  \bibinfo {author} {\bibfnamefont {J.}~\bibnamefont {Lenzner}}, \bibinfo
  {author} {\bibfnamefont {M.}~\bibnamefont {Lorenz}}, \bibinfo {author}
  {\bibfnamefont {H.}~\bibnamefont {Von~Wenckstern}}, \bibinfo {author}
  {\bibfnamefont {R.}~\bibnamefont {Schmidt-Grund}}, \ and\ \bibinfo {author}
  {\bibfnamefont {M.}~\bibnamefont {Grundmann}},\ }\href
  {https://aip.scitation.org/doi/full/10.1063/1.4915627} {\bibfield  {journal}
  {\bibinfo  {journal} {J. Appl. Phys.}\ }\textbf {\bibinfo {volume} {117}},\
  \bibinfo {pages} {125703} (\bibinfo {year} {2015})}\BibitemShut {NoStop}%
\bibitem [{\citenamefont {Lide}(2004)}]{lide2004crc}%
  \BibitemOpen
  \bibfield  {author} {\bibinfo {author} {\bibfnamefont {D.~R.}\ \bibnamefont
  {Lide}},\ }\href@noop {} {\emph {\bibinfo {title} {CRC handbook of chemistry
  and physics}}},\ Vol.~\bibinfo {volume} {85}\ (\bibinfo  {publisher} {CRC
  press},\ \bibinfo {year} {2004})\BibitemShut {NoStop}%
\bibitem [{\citenamefont {Zhou}\ and\ \citenamefont
  {Snyder}(1991)}]{zhou1991structures}%
  \BibitemOpen
  \bibfield  {author} {\bibinfo {author} {\bibfnamefont {R.-S.}\ \bibnamefont
  {Zhou}}\ and\ \bibinfo {author} {\bibfnamefont {R.~L.}\ \bibnamefont
  {Snyder}},\ }\href {https://scripts.iucr.org/cgi-bin/paper?st0485} {\bibfield
   {journal} {\bibinfo  {journal} {Acta Crystallogr. B}\ }\textbf {\bibinfo
  {volume} {47}},\ \bibinfo {pages} {617} (\bibinfo {year} {1991})}\BibitemShut
  {NoStop}%
\bibitem [{\citenamefont {Perdew}\ \emph {et~al.}(1996)\citenamefont {Perdew},
  \citenamefont {Burke},\ and\ \citenamefont {Ernzerhof}}]{Perdew1996}%
  \BibitemOpen
  \bibfield  {author} {\bibinfo {author} {\bibfnamefont {J.}~\bibnamefont
  {Perdew}}, \bibinfo {author} {\bibfnamefont {K.}~\bibnamefont {Burke}}, \
  and\ \bibinfo {author} {\bibfnamefont {M.}~\bibnamefont {Ernzerhof}},\ }\href
  {\doibase 10.1103/PhysRevLett.77.3865} {\bibfield  {journal} {\bibinfo
  {journal} {Phys. Rev. Lett.}\ }\textbf {\bibinfo {volume} {77}},\ \bibinfo
  {pages} {3865} (\bibinfo {year} {1996})}\BibitemShut {NoStop}%
\bibitem [{\citenamefont {Henkelman}\ \emph {et~al.}(2000)\citenamefont
  {Henkelman}, \citenamefont {Uberuaga},\ and\ \citenamefont
  {J{\'o}nsson}}]{henkelman2000climbing}%
  \BibitemOpen
  \bibfield  {author} {\bibinfo {author} {\bibfnamefont {G.}~\bibnamefont
  {Henkelman}}, \bibinfo {author} {\bibfnamefont {B.~P.}\ \bibnamefont
  {Uberuaga}}, \ and\ \bibinfo {author} {\bibfnamefont {H.}~\bibnamefont
  {J{\'o}nsson}},\ }\href {https://aip.scitation.org/doi/abs/10.1063/1.1329672}
  {\bibfield  {journal} {\bibinfo  {journal} {J. Chem. Phys.}\ }\textbf
  {\bibinfo {volume} {113}},\ \bibinfo {pages} {9901} (\bibinfo {year}
  {2000})}\BibitemShut {NoStop}%
\bibitem [{\citenamefont {Vineyard}(1957)}]{vineyard1957frequency}%
  \BibitemOpen
  \bibfield  {author} {\bibinfo {author} {\bibfnamefont {G.~H.}\ \bibnamefont
  {Vineyard}},\ }\href
  {https://www.sciencedirect.com/science/article/pii/0022369757900598}
  {\bibfield  {journal} {\bibinfo  {journal} {J. Phys. Chem. Solids}\ }\textbf
  {\bibinfo {volume} {3}},\ \bibinfo {pages} {121} (\bibinfo {year}
  {1957})}\BibitemShut {NoStop}%
\bibitem [{\citenamefont {Weiser}\ \emph {et~al.}(2018)\citenamefont {Weiser},
  \citenamefont {Stavola}, \citenamefont {Fowler}, \citenamefont {Qin},\ and\
  \citenamefont {Pearton}}]{weiser2018structure}%
  \BibitemOpen
  \bibfield  {author} {\bibinfo {author} {\bibfnamefont {P.}~\bibnamefont
  {Weiser}}, \bibinfo {author} {\bibfnamefont {M.}~\bibnamefont {Stavola}},
  \bibinfo {author} {\bibfnamefont {W.~B.}\ \bibnamefont {Fowler}}, \bibinfo
  {author} {\bibfnamefont {Y.}~\bibnamefont {Qin}}, \ and\ \bibinfo {author}
  {\bibfnamefont {S.}~\bibnamefont {Pearton}},\ }\href
  {https://aip.scitation.org/doi/abs/10.1063/1.5029921} {\bibfield  {journal}
  {\bibinfo  {journal} {Appl. Phys. Lett.}\ }\textbf {\bibinfo {volume}
  {112}},\ \bibinfo {pages} {232104} (\bibinfo {year} {2018})}\BibitemShut
  {NoStop}%
\bibitem [{\citenamefont {Janotti}\ and\ \citenamefont {Van~de
  Walle}(2007)}]{janotti2007native}%
  \BibitemOpen
  \bibfield  {author} {\bibinfo {author} {\bibfnamefont {A.}~\bibnamefont
  {Janotti}}\ and\ \bibinfo {author} {\bibfnamefont {C.~G.}\ \bibnamefont
  {Van~de Walle}},\ }\href
  {https://journals.aps.org/prb/abstract/10.1103/PhysRevB.76.165202} {\bibfield
   {journal} {\bibinfo  {journal} {Phys. Rev. B}\ }\textbf {\bibinfo {volume}
  {76}},\ \bibinfo {pages} {165202} (\bibinfo {year} {2007})}\BibitemShut
  {NoStop}%
\bibitem [{\citenamefont {Freysoldt}\ \emph {et~al.}(2009)\citenamefont
  {Freysoldt}, \citenamefont {Neugebauer},\ and\ \citenamefont {Van~de
  Walle}}]{freysoldt2009fully}%
  \BibitemOpen
  \bibfield  {author} {\bibinfo {author} {\bibfnamefont {C.}~\bibnamefont
  {Freysoldt}}, \bibinfo {author} {\bibfnamefont {J.}~\bibnamefont
  {Neugebauer}}, \ and\ \bibinfo {author} {\bibfnamefont {C.~G.}\ \bibnamefont
  {Van~de Walle}},\ }\href
  {https://journals.aps.org/prl/abstract/10.1103/PhysRevLett.102.016402}
  {\bibfield  {journal} {\bibinfo  {journal} {Phys. Rev. Lett.}\ }\textbf
  {\bibinfo {volume} {102}},\ \bibinfo {pages} {016402} (\bibinfo {year}
  {2009})}\BibitemShut {NoStop}%
\bibitem [{\citenamefont {Freysoldt}\ \emph {et~al.}(2011)\citenamefont
  {Freysoldt}, \citenamefont {Neugebauer},\ and\ \citenamefont {Van~de
  Walle}}]{freysoldt2011electrostatic}%
  \BibitemOpen
  \bibfield  {author} {\bibinfo {author} {\bibfnamefont {C.}~\bibnamefont
  {Freysoldt}}, \bibinfo {author} {\bibfnamefont {J.}~\bibnamefont
  {Neugebauer}}, \ and\ \bibinfo {author} {\bibfnamefont {C.~G.}\ \bibnamefont
  {Van~de Walle}},\ }\href
  {https://onlinelibrary.wiley.com/doi/abs/10.1002/pssb.201046289} {\bibfield
  {journal} {\bibinfo  {journal} {Phys. Status Solidi B}\ }\textbf {\bibinfo
  {volume} {248}},\ \bibinfo {pages} {1067} (\bibinfo {year}
  {2011})}\BibitemShut {NoStop}%
\bibitem [{\citenamefont {Lyons}\ \emph
  {et~al.}(2014{\natexlab{b}})\citenamefont {Lyons}, \citenamefont {Janotti},\
  and\ \citenamefont {Van~de Walle}}]{lyons2014cnitrides}%
  \BibitemOpen
  \bibfield  {author} {\bibinfo {author} {\bibfnamefont {J.~L.}\ \bibnamefont
  {Lyons}}, \bibinfo {author} {\bibfnamefont {A.}~\bibnamefont {Janotti}}, \
  and\ \bibinfo {author} {\bibfnamefont {C.~G.}\ \bibnamefont {Van~de Walle}},\
  }\href {https://journals.aps.org/prb/abstract/10.1103/PhysRevB.89.035204}
  {\bibfield  {journal} {\bibinfo  {journal} {Phys. Rev. B}\ }\textbf {\bibinfo
  {volume} {89}},\ \bibinfo {pages} {035204} (\bibinfo {year}
  {2014}{\natexlab{b}})}\BibitemShut {NoStop}%
\bibitem [{\citenamefont {Lyons}\ and\ \citenamefont {Van~de
  Walle}(2017)}]{lyons2017npjgan}%
  \BibitemOpen
  \bibfield  {author} {\bibinfo {author} {\bibfnamefont {J.~L.}\ \bibnamefont
  {Lyons}}\ and\ \bibinfo {author} {\bibfnamefont {C.~G.}\ \bibnamefont {Van~de
  Walle}},\ }\href {https://www.nature.com/articles/s41524-017-0014-2}
  {\bibfield  {journal} {\bibinfo  {journal} {npj Comput. Mater.}\ }\textbf
  {\bibinfo {volume} {3}},\ \bibinfo {pages} {1} (\bibinfo {year}
  {2017})}\BibitemShut {NoStop}%
\bibitem [{\citenamefont {McCluskey}\ \emph {et~al.}(1998)\citenamefont
  {McCluskey}, \citenamefont {Johnson}, \citenamefont {Van~de Walle},
  \citenamefont {Bour}, \citenamefont {Kneissl},\ and\ \citenamefont
  {Walukiewicz}}]{mccluskey1998metastability}%
  \BibitemOpen
  \bibfield  {author} {\bibinfo {author} {\bibfnamefont {M.}~\bibnamefont
  {McCluskey}}, \bibinfo {author} {\bibfnamefont {N.}~\bibnamefont {Johnson}},
  \bibinfo {author} {\bibfnamefont {C.~G.}\ \bibnamefont {Van~de Walle}},
  \bibinfo {author} {\bibfnamefont {D.}~\bibnamefont {Bour}}, \bibinfo {author}
  {\bibfnamefont {M.}~\bibnamefont {Kneissl}}, \ and\ \bibinfo {author}
  {\bibfnamefont {W.}~\bibnamefont {Walukiewicz}},\ }\href
  {https://journals.aps.org/prl/abstract/10.1103/PhysRevLett.80.4008}
  {\bibfield  {journal} {\bibinfo  {journal} {Phys. Rev. Lett.}\ }\textbf
  {\bibinfo {volume} {80}},\ \bibinfo {pages} {4008} (\bibinfo {year}
  {1998})}\BibitemShut {NoStop}%
\bibitem [{\citenamefont {Skierbiszewski}\ \emph {et~al.}(1999)\citenamefont
  {Skierbiszewski}, \citenamefont {Suski}, \citenamefont {Leszczynski},
  \citenamefont {Shin}, \citenamefont {Skowronski}, \citenamefont {Bremser},\
  and\ \citenamefont {Davis}}]{skierbiszewski1999evidence}%
  \BibitemOpen
  \bibfield  {author} {\bibinfo {author} {\bibfnamefont {C.}~\bibnamefont
  {Skierbiszewski}}, \bibinfo {author} {\bibfnamefont {T.}~\bibnamefont
  {Suski}}, \bibinfo {author} {\bibfnamefont {M.}~\bibnamefont {Leszczynski}},
  \bibinfo {author} {\bibfnamefont {M.}~\bibnamefont {Shin}}, \bibinfo {author}
  {\bibfnamefont {M.}~\bibnamefont {Skowronski}}, \bibinfo {author}
  {\bibfnamefont {M.}~\bibnamefont {Bremser}}, \ and\ \bibinfo {author}
  {\bibfnamefont {R.}~\bibnamefont {Davis}},\ }\href
  {https://aip.scitation.org/doi/abs/10.1063/1.124195} {\bibfield  {journal}
  {\bibinfo  {journal} {Appl. Phys. lett.}\ }\textbf {\bibinfo {volume} {74}},\
  \bibinfo {pages} {3833} (\bibinfo {year} {1999})}\BibitemShut {NoStop}%
\bibitem [{\citenamefont {Bouzid}\ and\ \citenamefont
  {Pasquarello}(2019)}]{bouzid2019defect}%
  \BibitemOpen
  \bibfield  {author} {\bibinfo {author} {\bibfnamefont {A.}~\bibnamefont
  {Bouzid}}\ and\ \bibinfo {author} {\bibfnamefont {A.}~\bibnamefont
  {Pasquarello}},\ }\href
  {https://onlinelibrary.wiley.com/doi/full/10.1002/pssr.201800633} {\bibfield
  {journal} {\bibinfo  {journal} {Phys. Status Solidi RRL}\ }\textbf {\bibinfo
  {volume} {13}},\ \bibinfo {pages} {1800633} (\bibinfo {year}
  {2019})}\BibitemShut {NoStop}%
\bibitem [{\citenamefont {Choi}\ \emph {et~al.}(2013)\citenamefont {Choi},
  \citenamefont {Lyons}, \citenamefont {Janotti},\ and\ \citenamefont {Van~de
  Walle}}]{choi2013impact}%
  \BibitemOpen
  \bibfield  {author} {\bibinfo {author} {\bibfnamefont {M.}~\bibnamefont
  {Choi}}, \bibinfo {author} {\bibfnamefont {J.~L.}\ \bibnamefont {Lyons}},
  \bibinfo {author} {\bibfnamefont {A.}~\bibnamefont {Janotti}}, \ and\
  \bibinfo {author} {\bibfnamefont {C.~G.}\ \bibnamefont {Van~de Walle}},\
  }\href {https://aip.scitation.org/doi/abs/10.1063/1.4801497} {\bibfield
  {journal} {\bibinfo  {journal} {Appl. Phys. Lett.}\ }\textbf {\bibinfo
  {volume} {102}},\ \bibinfo {pages} {142902} (\bibinfo {year}
  {2013})}\BibitemShut {NoStop}%
\bibitem [{\citenamefont {Van~de Walle}\ and\ \citenamefont
  {Neugebauer}(2003)}]{van2003universal}%
  \BibitemOpen
  \bibfield  {author} {\bibinfo {author} {\bibfnamefont {C.~G.}\ \bibnamefont
  {Van~de Walle}}\ and\ \bibinfo {author} {\bibfnamefont {J.}~\bibnamefont
  {Neugebauer}},\ }\href {https://www.nature.com/articles/nature01665}
  {\bibfield  {journal} {\bibinfo  {journal} {Nature}\ }\textbf {\bibinfo
  {volume} {423}},\ \bibinfo {pages} {626} (\bibinfo {year}
  {2003})}\BibitemShut {NoStop}%
\bibitem [{\citenamefont {Mu}\ \emph {et~al.}(2020)\citenamefont {Mu},
  \citenamefont {Peelaers}, \citenamefont {Zhang}, \citenamefont {Wang},\ and\
  \citenamefont {Van~de Walle}}]{mu2020orientation}%
  \BibitemOpen
  \bibfield  {author} {\bibinfo {author} {\bibfnamefont {S.}~\bibnamefont
  {Mu}}, \bibinfo {author} {\bibfnamefont {H.}~\bibnamefont {Peelaers}},
  \bibinfo {author} {\bibfnamefont {Y.}~\bibnamefont {Zhang}}, \bibinfo
  {author} {\bibfnamefont {M.}~\bibnamefont {Wang}}, \ and\ \bibinfo {author}
  {\bibfnamefont {C.~G.}\ \bibnamefont {Van~de Walle}},\ }\href
  {https://aip.scitation.org/doi/abs/10.1063/5.0036072} {\bibfield  {journal}
  {\bibinfo  {journal} {Appl. Phys. Lett.}\ }\textbf {\bibinfo {volume}
  {117}},\ \bibinfo {pages} {252104} (\bibinfo {year} {2020})}\BibitemShut
  {NoStop}%
\bibitem [{\citenamefont {Van~de Walle}\ and\ \citenamefont
  {Neugebauer}(2004)}]{jap04}%
  \BibitemOpen
  \bibfield  {author} {\bibinfo {author} {\bibfnamefont {C.~G.}\ \bibnamefont
  {Van~de Walle}}\ and\ \bibinfo {author} {\bibfnamefont {J.}~\bibnamefont
  {Neugebauer}},\ }\href {https://aip.scitation.org/doi/abs/10.1063/1.1682673}
  {\bibfield  {journal} {\bibinfo  {journal} {J. Appl. Phys.}\ }\textbf
  {\bibinfo {volume} {95}},\ \bibinfo {pages} {3851} (\bibinfo {year}
  {2004})}\BibitemShut {NoStop}%
\bibitem [{\citenamefont {Venzie}\ \emph {et~al.}(2021)\citenamefont {Venzie},
  \citenamefont {Portoff}, \citenamefont {Fares}, \citenamefont {Stavola},
  \citenamefont {Fowler}, \citenamefont {Ren},\ and\ \citenamefont
  {Pearton}}]{venzie2021oh}%
  \BibitemOpen
  \bibfield  {author} {\bibinfo {author} {\bibfnamefont {A.}~\bibnamefont
  {Venzie}}, \bibinfo {author} {\bibfnamefont {A.}~\bibnamefont {Portoff}},
  \bibinfo {author} {\bibfnamefont {C.}~\bibnamefont {Fares}}, \bibinfo
  {author} {\bibfnamefont {M.}~\bibnamefont {Stavola}}, \bibinfo {author}
  {\bibfnamefont {W.~B.}\ \bibnamefont {Fowler}}, \bibinfo {author}
  {\bibfnamefont {F.}~\bibnamefont {Ren}}, \ and\ \bibinfo {author}
  {\bibfnamefont {S.~J.}\ \bibnamefont {Pearton}},\ }\href
  {https://aip.scitation.org/doi/abs/10.1063/5.0059769} {\bibfield  {journal}
  {\bibinfo  {journal} {Appl. Phys. Lett.}\ }\textbf {\bibinfo {volume}
  {119}},\ \bibinfo {pages} {062109} (\bibinfo {year} {2021})}\BibitemShut
  {NoStop}%
\bibitem [{\citenamefont {Van~de Walle}\ and\ \citenamefont
  {Neugebauer}(2006)}]{AR2006}%
  \BibitemOpen
  \bibfield  {author} {\bibinfo {author} {\bibfnamefont {C.~G.}\ \bibnamefont
  {Van~de Walle}}\ and\ \bibinfo {author} {\bibfnamefont {J.}~\bibnamefont
  {Neugebauer}},\ }\href
  {https://www.annualreviews.org/doi/abs/10.1146/annurev.matsci.36.010705.155428}
  {\bibfield  {journal} {\bibinfo  {journal} {Annu. Rev. Mater. Res.}\ }\textbf
  {\bibinfo {volume} {36}},\ \bibinfo {pages} {179} (\bibinfo {year}
  {2006})}\BibitemShut {NoStop}%
\bibitem [{SM()}]{SM}%
  \BibitemOpen
  \href@noop {} {}\bibinfo {note} {See Supplemental Material at [URL will be
  inserted by publisher] for the formation energy of Si$_\text{O}$ and
  Si$_\text{O}$--H complexes in Ga$_2$O$_3$ and Al$_2$O$_3$.}\BibitemShut
  {Stop}%
\bibitem [{\citenamefont {Choi}\ \emph {et~al.}(2014)\citenamefont {Choi},
  \citenamefont {Janotti},\ and\ \citenamefont {Van~de
  Walle}}]{choi2014hydrogen}%
  \BibitemOpen
  \bibfield  {author} {\bibinfo {author} {\bibfnamefont {M.}~\bibnamefont
  {Choi}}, \bibinfo {author} {\bibfnamefont {A.}~\bibnamefont {Janotti}}, \
  and\ \bibinfo {author} {\bibfnamefont {C.~G.}\ \bibnamefont {Van~de Walle}},\
  }\href {https://pubs.acs.org/doi/abs/10.1021/am4057997} {\bibfield  {journal}
  {\bibinfo  {journal} {ACS Appl. Mater. Interfaces}\ }\textbf {\bibinfo
  {volume} {6}},\ \bibinfo {pages} {4149} (\bibinfo {year} {2014})}\BibitemShut
  {NoStop}%
\bibitem [{\citenamefont {Peelaers}\ \emph
  {et~al.}(2019{\natexlab{b}})\citenamefont {Peelaers}, \citenamefont {Lyons},
  \citenamefont {Varley},\ and\ \citenamefont {Van~de
  Walle}}]{peelaers2019deep}%
  \BibitemOpen
  \bibfield  {author} {\bibinfo {author} {\bibfnamefont {H.}~\bibnamefont
  {Peelaers}}, \bibinfo {author} {\bibfnamefont {J.~L.}\ \bibnamefont {Lyons}},
  \bibinfo {author} {\bibfnamefont {J.~B.}\ \bibnamefont {Varley}}, \ and\
  \bibinfo {author} {\bibfnamefont {C.~G.}\ \bibnamefont {Van~de Walle}},\
  }\href {https://aip.scitation.org/doi/abs/10.1063/1.5063807} {\bibfield
  {journal} {\bibinfo  {journal} {APL Mater.}\ }\textbf {\bibinfo {volume}
  {7}},\ \bibinfo {pages} {022519} (\bibinfo {year}
  {2019}{\natexlab{b}})}\BibitemShut {NoStop}%
\bibitem [{\citenamefont {Lany}(2018)}]{lany2018defect}%
  \BibitemOpen
  \bibfield  {author} {\bibinfo {author} {\bibfnamefont {S.}~\bibnamefont
  {Lany}},\ }\href {https://aip.scitation.org/doi/10.1063/1.5019938} {\bibfield
   {journal} {\bibinfo  {journal} {APL Mater.}\ }\textbf {\bibinfo {volume}
  {6}},\ \bibinfo {pages} {046103} (\bibinfo {year} {2018})}\BibitemShut
  {NoStop}%
\bibitem [{\citenamefont {Baker}\ \emph {et~al.}(2020)\citenamefont {Baker},
  \citenamefont {Bowes}, \citenamefont {Harris}, \citenamefont {Collazo},
  \citenamefont {Sitar},\ and\ \citenamefont {Irving}}]{Baker2020}%
  \BibitemOpen
  \bibfield  {author} {\bibinfo {author} {\bibfnamefont {J.~N.}\ \bibnamefont
  {Baker}}, \bibinfo {author} {\bibfnamefont {P.~C.}\ \bibnamefont {Bowes}},
  \bibinfo {author} {\bibfnamefont {J.~S.}\ \bibnamefont {Harris}}, \bibinfo
  {author} {\bibfnamefont {R.}~\bibnamefont {Collazo}}, \bibinfo {author}
  {\bibfnamefont {Z.}~\bibnamefont {Sitar}}, \ and\ \bibinfo {author}
  {\bibfnamefont {D.~L.}\ \bibnamefont {Irving}},\ }\href {\doibase
  10.1063/5.0013988} {\bibfield  {journal} {\bibinfo  {journal} {Appl. Phys.
  Lett.}\ }\textbf {\bibinfo {volume} {117}},\ \bibinfo {pages} {102109}
  (\bibinfo {year} {2020})}\BibitemShut {NoStop}%
\bibitem [{\citenamefont {Bogus{\l}awski}\ and\ \citenamefont
  {Bernholc}(1997)}]{boguslawski1997doping}%
  \BibitemOpen
  \bibfield  {author} {\bibinfo {author} {\bibfnamefont {P.}~\bibnamefont
  {Bogus{\l}awski}}\ and\ \bibinfo {author} {\bibfnamefont {J.}~\bibnamefont
  {Bernholc}},\ }\href
  {https://journals.aps.org/prb/abstract/10.1103/PhysRevB.56.9496} {\bibfield
  {journal} {\bibinfo  {journal} {Phys. Rev. B}\ }\textbf {\bibinfo {volume}
  {56}},\ \bibinfo {pages} {9496} (\bibinfo {year} {1997})}\BibitemShut
  {NoStop}%
\end{thebibliography}
\end{document}